\documentclass[aip,jcp,preprint,author-numerical]{revtex4-2}
\usepackage{balance}
\usepackage{dcolumn}
\usepackage{float}
\usepackage{array}
\usepackage[utf8]{inputenc}
\usepackage[T1]{fontenc}
\usepackage{mathptmx}
\usepackage{etoolbox}
\usepackage[version=3]{mhchem}        
\usepackage{graphicx}                
\usepackage{textcomp, gensymb}       
\usepackage[export]{adjustbox}       
\usepackage{amsmath, amssymb, amsfonts} 
\usepackage{mathtools}               
\usepackage{booktabs, tabularx, ragged2e} 
\def\@email#1#2{%
 \endgroup
 \patchcmd{\titleblock@produce}
  {\frontmatter@RRAPformat}
  {\frontmatter@RRAPformat{\produce@RRAP{*#1\href{mailto:#2}{#2}}}\frontmatter@RRAPformat}
  {}{}
}%
\makeatother
\begin{document}
\title{Infrared Spectral Signature of Water as a Probe to Demystify Urea Aggregation and Force Field Accuracy}
\author{Pankaj Adhikary}
\author{Rajib Biswas}
\affiliation{Department of Chemistry, Indian Institute of Technology Tirupati, Tirupati, India}
\email{rajib@iittp.ac.in}
\date{\today}

\begin{abstract}
Urea is a widely used as a protein denaturant; however, the potential of urea to form self-assembled structures at higher concentrations and the influence of its self-interactions on water structure and dynamics remains elusive. This open question demands tracking of molecular-level rearrangements. In this work, we explore the influence of urea on water's local structure and dynamics and relate it to urea self-association. We correlate vibrational spectral response and orientational dynamics of water with concentration-dependent self-association of urea by looking at the interface surface area, hydrogen bond strength, and population of relevant donor-acceptor pairs. We compare the response of four urea force fields (KBFF, OPLS-S, OPLS-AA-D, GAFF-D3) with SPC/E water. The KBFF–SPC/E pair best reproduces experimental IR spectra. Both variants of the Duffy model (OPLS-S, OPLS-AA-D) show blue shifts with reasonable broadening and strong concentration-dependent responses, while GAFF-D3 shows random peak shifts with prominent broadening. Regarding urea self-aggregation, KBFF is mildly repulsive, Duffy models are attractive, and GAFF-D3 is neutral with high variability. Only KBFF–SPC/E and GAFF-D3–SPC/E pairs capture the expected deceleration in water-orientational dynamics. We conclude that urea does not self-aggregate significantly in water, even at higher concentrations. KBFF emerges as the most reliable classical non-polarizable model of urea for capturing both structural and dynamic properties of water.
\end{abstract}
\maketitle
\section{\label{sec:level1}Introduction}
Urea is a pervasive denaturing agent whose microscopic mechanism remains debated,\cite{Schellman2002,kresheck1965temperature,Bennion2003,Das2009} whether via direct interaction with the polypeptides or indirect solvent-mediated 
perturbation. Consequently, it received plenty of interest and remains a focus of active research. There are mainly two schools of thought for the urea-induced protein denaturation mechanism: indirect and direct interaction. The indirect mechanism proposes that urea reduces the hydrophobic effect by disrupting the water structure and enhancing solvation of hydrophobic residues, thereby destabilizing folded proteins.\cite{Bennion2003,rezus2006effect,Rezus2007,Sagle2009} Conversely, the direct interaction mechanism proposes that urea directly forms hydrogen bonds with the peptide backbone, ultimately favouring the unfolded state, as evidenced by recent molecular dynamics simulations highlighting urea’s preferential binding to nonpolar and polar residues. While early studies emphasized indirect effects, modern simulations increasingly support a combined mechanism, with direct interactions playing a dominant role in unfolding.\cite{Rosgen2000,Zou2002,Auton2004,stumpe2007aqueous,Auton2007,das2008atomistic,Stumpe2008,carr2013structure,Candotti2013,roy2014comparative,Arsiccio2022,ADHIKARY2024100609} The microscopic understanding of urea-driven protein denaturation is complicated by the dynamic interplay among proteins, water, and urea in ternary systems. A molecular level understanding requires a qualitative and structural grasp of the intricate balance between intra-protein, protein -- solvent/co-solvent, and co-solvent -- co-solvent/solvent interactions. This work focuses specifically on resolving the role of co-solvent -- solvent and intra co-solvent interactions. 
\par 
There are multiple experimental and simulation studies focusing on co-solvent -- solvent interaction, indicating that urea is inert towards water structure and minimally perturbs the hydrogen bond (H-bond) network. The OD and OH stretching bands in the IR spectra remain unaltered even up to high urea concentrations,\cite{carr2013structure,ojha2019urea,ADHIKARY2024100609} suggesting urea barely disrupts water’s H-bond network. Molecular dynamics (MD) simulations concur: urea has the right size to fit into water’s H-bonded network without destroying the tetrahedral arrangement.\cite{bandyopadhyay2014molecular} Extensive spectroscopic and simulation studies over the past few decades converged on the conclusion that the addition of urea to water causes a subtle yet evident slowdown of the second-order Legendre correlation function C$_2$(t), reflecting hindered rotational dynamics of water molecules. The time‐resolved experiments by Rezus and Bakker demonstrate that even modest urea concentrations introduce a slower decay component in C$_2$(t), indicative of immobilized water populations. Pump–probe and dielectric relaxation measurements further reveal a bimodal reorientation behaviour in urea–water mixtures: a majority of water molecules retain bulk‐like reorientation times (~2.5–2.7 ps), while a minority fraction exhibits dramatically prolonged orientational lifetimes that increase linearly with urea concentration.\cite{rezus2006effect, agieienko2016urea} However, MD simulations suggest a more homogeneous slowdown arising from steric hindrance by urea's excluded volume that blocks angular jumps of nearby waters.\cite{laage2011reorientation,carr2013structure} 
\par
\begin{table*}[hbtp!]
    \centering
    \caption{Systems used for classical MD simulation.}
    \begin{tabular}{|c|c|c|c|c|}
        \hline
        \textbf{Nominal Urea Conc. (M)} & \textbf{Actual Urea Conc. (M)} & \textbf{$\mathrm{N_{Urea}}$} & \textbf{$\mathrm{{N_Water}}$} & \textbf{Box Size (nm)} \\
        \hline
        0  & 0.00  & 0   & 2213 & 4.0494 \\
        \hline
        1  & 1.03  & 41  & 2115 & 4.0469 \\
        \hline
        4  & 4.07  & 161 & 1818 & 4.0507 \\
        \hline
        6  & 6.09  & 241 & 1620 & 4.0593 \\
        \hline
        8  & 8.23  & 321 & 1420 & 4.0368 \\
        \hline
        11 & 11.33 & 441 & 1221 & 4.0383 \\
        \hline
    \end{tabular}
    \label{tab:system}
\end{table*}
Microscopic understanding of intra-urea interaction is crucial because it provides comparative insights into intra-urea and urea-water interactions, which can directly impact its denaturation efficiency and the structural integrity of proteins in solution.\cite{Chen2016,Arsiccio2022,Yang2024} While explanations for urea's behaviour in water have varied, one consistent theme is the importance of hydrogen bonding. In water, urea shows an amphiphilic nature: its oxygen and hydrogen atoms form hydrogen bonds. In contrast, its carbon and nitrogen atoms are incapable of strong, short-range, noncovalent interactions and behave as hydrophobic moieties.\cite{vanzi1998effect,mountain2004importance} Research into the H-bonding energies between urea and water reveals that water-water (W-W) bonds are considerably stronger than those between urea-water (U-W) or between urea molecules themselves (U-U). This finding indicates that water molecules tend to maintain their own molecular network, leaving urea molecules to cluster together.\cite{stumpe2007aqueous} However, in our previous work,\cite{ADHIKARY2024100609} we find that U-U and U-W hydrogen bonding energies are comparable, and U-W H-bond is a little more stable than U-U H-bond. Urea's behaviour changes predictably with varying temperatures and concentrations. As temperature increases, urea's aggregation in water decreases, corresponding to higher solubility.\cite{weerasinghe2003kirkwood} In earlier work, it was also shown that the interactions between U-W and U-U reduce as temperature rises.\cite{mountain2004importance}
\par 
However, previous studies present disagreeing conclusions, complicating the interpretation of concentration-dependent trends. Grdadolnik and Maréchal \cite{grdadolnik2002urea} reported that U-U interactions occur above 1 M and grow linearly with concentrations up to 11 M and found 20\% of the urea molecules are primarily surrounded by other urea molecules; Stumpe et al. \cite{stumpe2007aqueous} suggested that aggregation starts around 6 M. In a similar line, Jung et al. \cite{jung2004characterization} provide a more detailed picture of successive stages of urea aggregation from the ribbon (linear), chain (cyclic) dimers to oligomers, and, finally, polymeric chains upon increasing urea concentration from 0.5 M to 6 M. This theory is further supported by previous studies.\cite{mafy2015effect,atahar2019aggregation} In contrast, other simulations report only small urea clusters.\cite{kokubo2007molecular} Funkner et al. \cite{funkner2012urea} observed no self-association up to 10 M.
These consequences imply that precise force fields must neither grossly overestimate urea self-association nor falsely suppress it, since both scenarios would mislead interpretations of urea’s action on water and proteins. Whether these opposing findings result from differences in the choice of force fields in simulation studies or interpretation, or are pieces of a larger, unresolved riddle is still an open question. The proliferation of various urea force fields created uncertainty in selecting the optimal model for simulating urea in aqueous environments. Although many parameter sets have been designed to reproduce gas, solution, and crystalline phase properties, their performance in water often diverges, impacting predicted structural, dynamic, and thermodynamic behaviour.\cite{duffy@1993,weerasinghe2003kirkwood,smith2004computer,ozpinar2010improved, anker2023assessment} Thus, a unified, systematic evaluation of leading urea force fields under identical aqueous conditions is essential to guide model choice and enhance simulation fidelity.
\\
\indent In this work, we examine four popular all‑atom urea models – a Kirkwood–Buff–fitted urea force field (KBFF),\cite{weerasinghe2003kirkwood} the OPLS‑S model,\cite{smith2004computer} an OPLS‑AA variant using Duffy’s urea charges(OPLS-AA-D),\cite{duffy@1993,anker2023assessment} and a GAFF‑based model with RESP charges (GAFF‑D3).\cite{ozpinar2010improved} Prior literature indicates that the KB–derived model reproduces ideal‐solution thermodynamics and bulk densities and diffusivities better than standard OPLS. At the same time, OPLS‑S has been widely used for urea crystallization simulations.\cite{weerasinghe2003kirkwood,kokubo2007molecular,piana2005understanding} In contrast, OPLS‑AA with Duffy charges tends to yield powerful cohesion (and high density) in urea crystals,\cite{das2008atomistic, Das2009} whereas GAFF‑D3 has been shown to capture urea’s crystal lattice and sublimation properties accurately.\cite{anker2023assessment} We employ theoretical vibrational spectroscopy to assess perturbations in water structure, compute potentials of mean force for urea–urea and urea–water pairs to measure intermolecular affinities, analyze cluster interface areas and urea–urea versus urea–water hydrogen bonds to quantify self‐aggregation, and calculate the second‐order Legendre correlation C$_2$(t) of water reorientation to evaluate rotational dynamics. Finally, H-bond strength ($\Delta G_{HB}$) provides the microscopic connection between urea-water and intra-urea interactions. These measures aim to refine the molecular level understanding of concentration-dependent self association of urea and its impact on water's local structure and dynamics. We also identify the efficacy of the standard classical non-polarizable urea force fields that best reproduce experimental and computational outcomes. We further cross-verify various contrasting interpretations from computational and experimental studies, thereby providing a robust foundation for urea’s effect on water structure and dynamics. The rest of this article is structured as follows. In the subsequent section, Sec. \ref{sec:level2}, we briefly describe the simulation details of classical molecular dynamics (MD), mixed quantum-classical approach for spectroscopy modeling, interface surface area, and H-bond configurational analysis methodology. Data analysis and significant observations are discussed in Sec. \ref{sec:level3}, and our conclusions are given in Sec. \ref{sec:level4}.

\section{\label{sec:level2}Methodology}
\subsection{\label{sec:level21}Classical molecular dynamics}
We perform classical MD simulations of bulk water and various concentrations of aqueous urea solution ranging from 1 M to 11 M as mentioned in Table \ref{tab:system} using the Groningen Machine for Chemical Simulations (GROMACS) version 2022.3 \cite{ABRAHAM201519}. The Extended Simple Point Charge (SPC/E) \cite{hjc1987grigera} parameters are used to model the bulk water, and four different force field parameters are used to describe urea, namely Kirkwood-Buff derived force field (KBFF);\cite{weerasinghe2003kirkwood} two variants of the Duffy \cite{duffy@1993} model (OPLS-S),\cite{smith2004computer} (OPLS-AA-D);\cite{duffy@1993,anker2023assessment} and Generalized Amber Force Field with RESP-D3 charges (GAFF-D3).\cite{ozpinar2010improved} OPLS-S and OPLS-AA-D have the same inter-molecular parameters of the Duffy\cite{duffy@1993} model and only differ by intra-molecular parameters. Initial configurations were constructed by randomly putting urea molecules first, then adding water molecules in a cubic box of side length 4.05 nm using the insert-molecules utility of GROMACS. Owing to their typically high potential energy, these initial configurations are subjected to energy minimization using the steepest descent algorithm, followed by a 500 ps NVT pre-equilibration run at 300 K with a time step of 1 fs using the V-rescale thermostat. Systems are further equilibrated in the NPT ensemble for 5 ns using a 1 fs time step. Finally, we perform a 1 ns production run using the NVT ensemble to extract the trajectory for analysis. We use Nos\'{e}-Hoover \cite{evans1985nose} thermostat with a relaxation time of 0.5 ps to maintain the temperature at 300 K. Parinello-Rahman \cite{parrinello1981polymorphic} barostat with a relaxation time of 1.0 ps preserves the pressure at 1 bar. A  15 \text{\AA} cutoff radius is used for neighbour searching and non-bonded interactions. All bonds are constrained using the LINCS algorithm.\cite{hess1997lincs} The long-range electrostatic interactions are maintained using the particle mesh Ewald (PME) \cite{darden1993particle} technique with a 1.6 \text{\AA} FFT grid spacing and an interpolation order of 4. We perform three independent sets of simulations with the mentioned protocol using different random seeds for each system for statistical robustness and reproducibility.

\subsection{\label{sec:level22}Mixed Quantum-Classical Simulation}
The mixed quantum-classical (MQC) technique is deployed to model the isotope dilute single-oscillator O-H stretch spectroscopy of the pure water and urea-water mixture. This is a reduced-dimensional semi-empirical mapping method widely employed earlier for various aqueous systems.\cite{corcelli2005infrared,Auer2008,SCHMIDT2007143,carr2013structure,reddy2020theoretical,reddy2023theoretical,ADHIKARY2024100609,Lin2009NaBr} Electronic structure calculations are performed on instantaneous small frozen clusters extracted from the classical molecular dynamics trajectory to construct the spectroscopic maps. These clusters portray a collection of samples of the local molecular surroundings of the solute in solution. This methodology facilitates the establishment of a correlation between the quantum mechanical O–H vibrational transition frequency and its associated transition dipole moment with a collective variable (CV). The CV is selected for its computational accessibility from classical molecular dynamics trajectories. This classical analogue is subsequently employed to construct time-dependent trajectories of both the transition frequency and dipole moment, which serve as critical inputs for spectral calculations based on a time-domain response function.
We leverage our previous scheme \cite{ADHIKARY2024100609} to select the instantaneous clusters from MD trajectories. Urea–water clusters are identified by selecting a central water oxygen atom ($O_w$) that is closest to the urea oxygen ($O_u$) in the carbonyl (-CO) ensemble or to the urea nitrogen ($N_u$) in the amine (-NH) ensemble. All molecules having their center of mass within a 7.0 \text{\AA} cutoff of the chosen central $O_w$ are included in the cluster. A detailed description of the cluster selection criteria is provided in the supporting information (section \textbf{S1}). We chose 200 urea-water clusters in total, consisting of 100 clusters of each type (CO, NH) for each spectroscopy map. In addition, we select 100 bulk water clusters. Each cluster includes $\sim 50$ water molecules, on average, adequate to create a bulk-like surroundings around the O–H bond of interest. Minor changes in cluster size do not affect the results, and the selected clusters thoroughly sample all statistically relevant configurations.\par
We compute the O–H stretching response for each chosen cluster based on the one-dimensional adiabatic potential energy surface (PES). The central O-H oscillator (\textit{r\textsubscript{OH}}) is elongated from 0.7 to 1.6 \text{\AA} with a uniform spacing of 0.08 \text{\AA} while keeping all other degrees of freedom fixed to obtain the quantum mechanical PE. The GAUSSIAN 16 package is employed for the calculation of single-point energy using DFT, utilizing the B3LYP hybrid functional \cite{lee1988development} and 6-311++G(d,p) basis set.\cite{frisch2016gaussian} The choice of the hybrid functional and the basis set is guided by their proven accuracy in computing potential energy along the O–H stretch coordinate, as documented in previous studies.\cite{corcelli2005infrared,skinner2013JCTC,biswas2016molecular,reddy2020theoretical,reddy2023theoretical} We produce spectroscopy maps using the three different force fields of urea: KBFF, OPLS-S, GAFF-D3 and SPC/E water and generate the spectra (see \textbf{SIV}). As the inter-molecular parameters are the same for both OPLS-AA-D and OPLS-S, therefore, we use the OPLS-S - SPC/E maps for the OPLS-AA-D - SPC/E pair.
\subsection{\label{sec:level23}Interface Surface Area}
To analyze and quantify the self-aggregation of urea molecules, We evaluate the decrease in total water-accessible surface area of urea relative to that in dilute urea in a similar manner to that of Stumpe et al.\cite{stumpe2007aqueous} The minimization of the interface surface area is presumed to be the main driving force for the accumulation. To differentiate ``geometric aggregation," which is a consequence of random contacts between urea molecules, from real aggregation, we have performed two extra types of simulations on all the urea water solutions. The first set of simulations involved using uncharged urea molecules to enhance the hydrophobic effect and promote maximum aggregation. The calculated surface area is used to define the 100\% aggregation level. In the succeeding set of simulations, both water and urea molecules are kept uncharged to eliminate any hydrophobic influences in the simulation. This approach led to the extreme scenario of purely stochastic clustering, defining the 0\% aggregation level. Three independent sets of simulations are done for each system, and for each case, i.e., for a particular molarity, 9 sets of simulations are done. These simulations were conducted under NVT ensemble to prevent the evaporation of the resulting Van der Waals liquid.

\subsection{\label{sec:level24}Hydrogen Bond Strength and Population}
To quantify the hydrogen bond (H-bond) strength ($\Delta G$), we follow a similar methodology developed by Sapir et al.\cite{sapir2017revisiting} We describe an H-bond between hydrogen (\textit{H}) linked to an electronegative atom ``donor" (\textit{D}) and an oxygen ``acceptor" (\textit{A}) atom by the \textit{D-A} distance, and the angle $\angle H-D-A$. $\Delta G_{HB}$ is described as the essential reversible work to shape the H-bonded probability distribution $P(r, \theta)$ from a random distribution $P_{\text{rand}}(r,\theta)$. We get the hydrogen bond strength by considering the entire configurational ensemble of donor-acceptor pairs in the first solvation shell. The $\Delta G_{HB}$ is calculated from  
\begin{equation}
\Delta G = -RT \int\limits_{0}^{\pi} \int\limits_{r_{min}}^{r_{max}} \zeta P(r,\theta) \ln\frac{\zeta P(r, \theta)}{\zeta_{\text{rand}} P_{\text{rand}}(r,\theta)} \, d\theta \, dr \label{delG_eqn},
\end{equation}
where R describes the gas constant, and T represents the absolute temperature. In our calculation, we consider T = 300 K. We provide a more detailed description of this process in the supporting information (section \textbf{S5}). For the H-bond population, we consider \textit{D-A} distance cutoff of 4.5~\AA{} for N$_u$-O$_W$ for the rest 3.5~\AA{} and an angle $\angle H-D-A$ cutoff of $30^\circ$ for all.\cite{luzar1996hydrogen}

\begin{figure*}[htbp!]
    \centering
    \includegraphics[width=0.45\textwidth, clip]{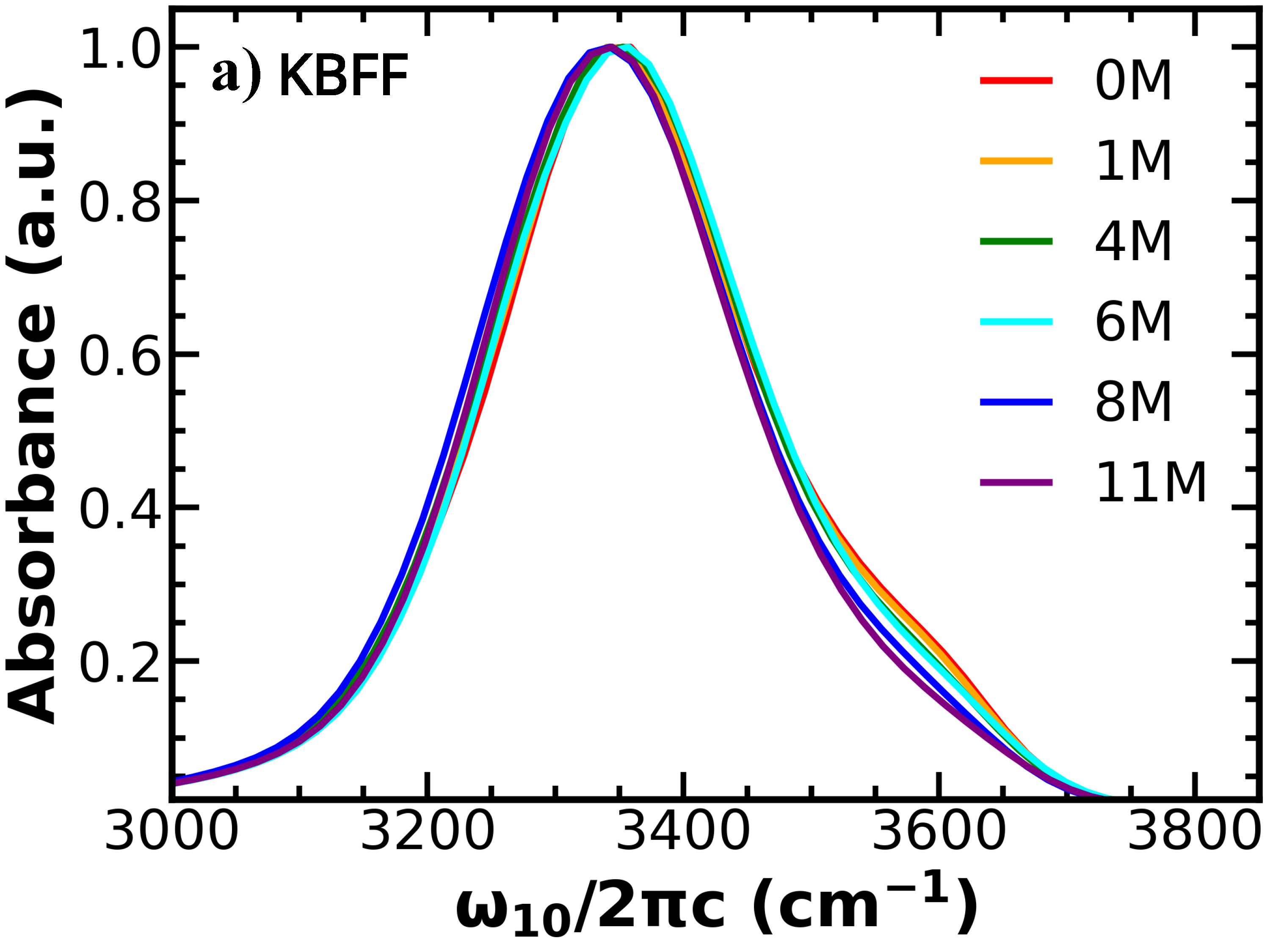} 
    \includegraphics[width=0.45\textwidth, clip]{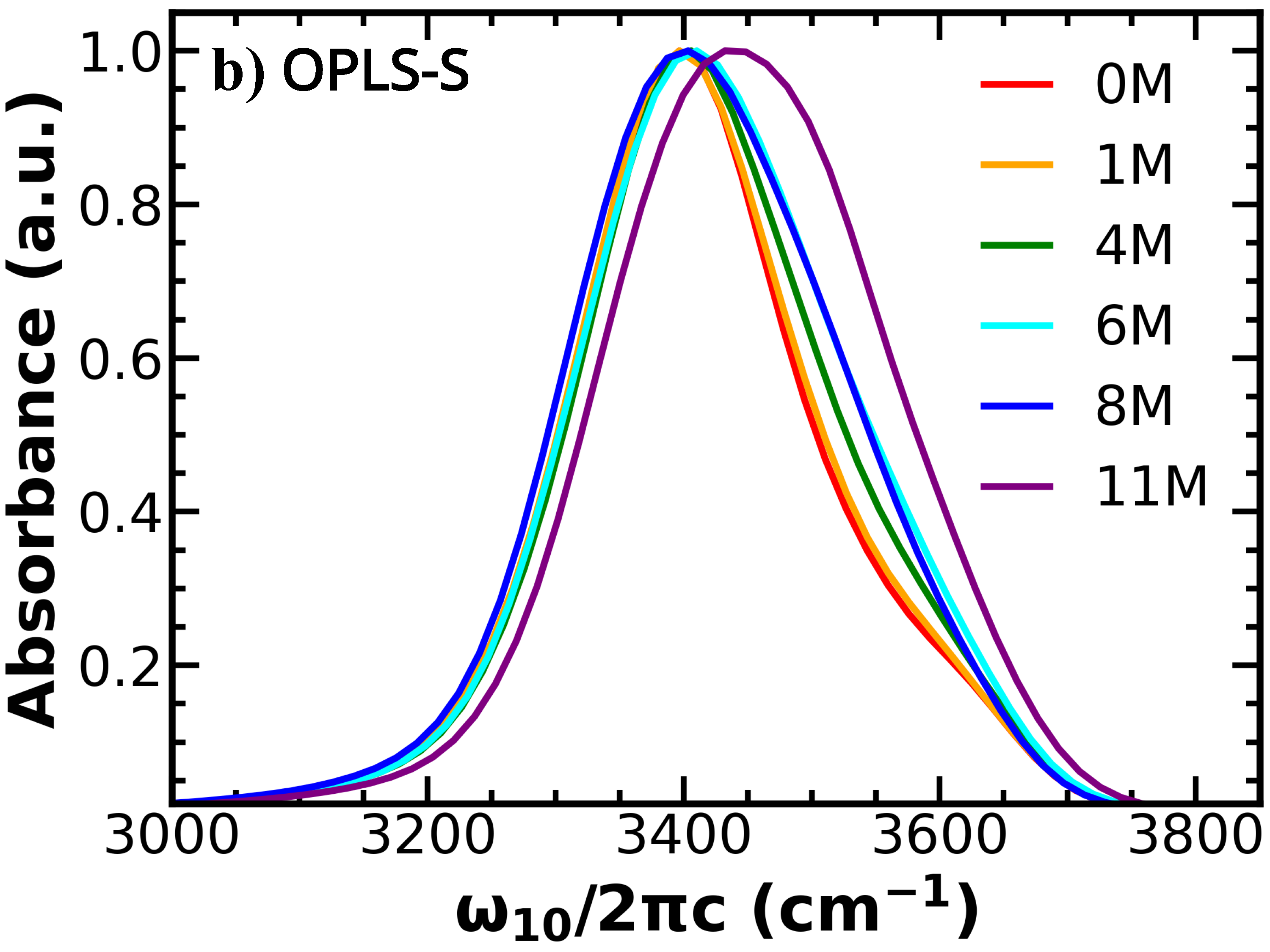}
        \vspace{0.25cm}  
    \includegraphics[width=0.45\textwidth, clip]{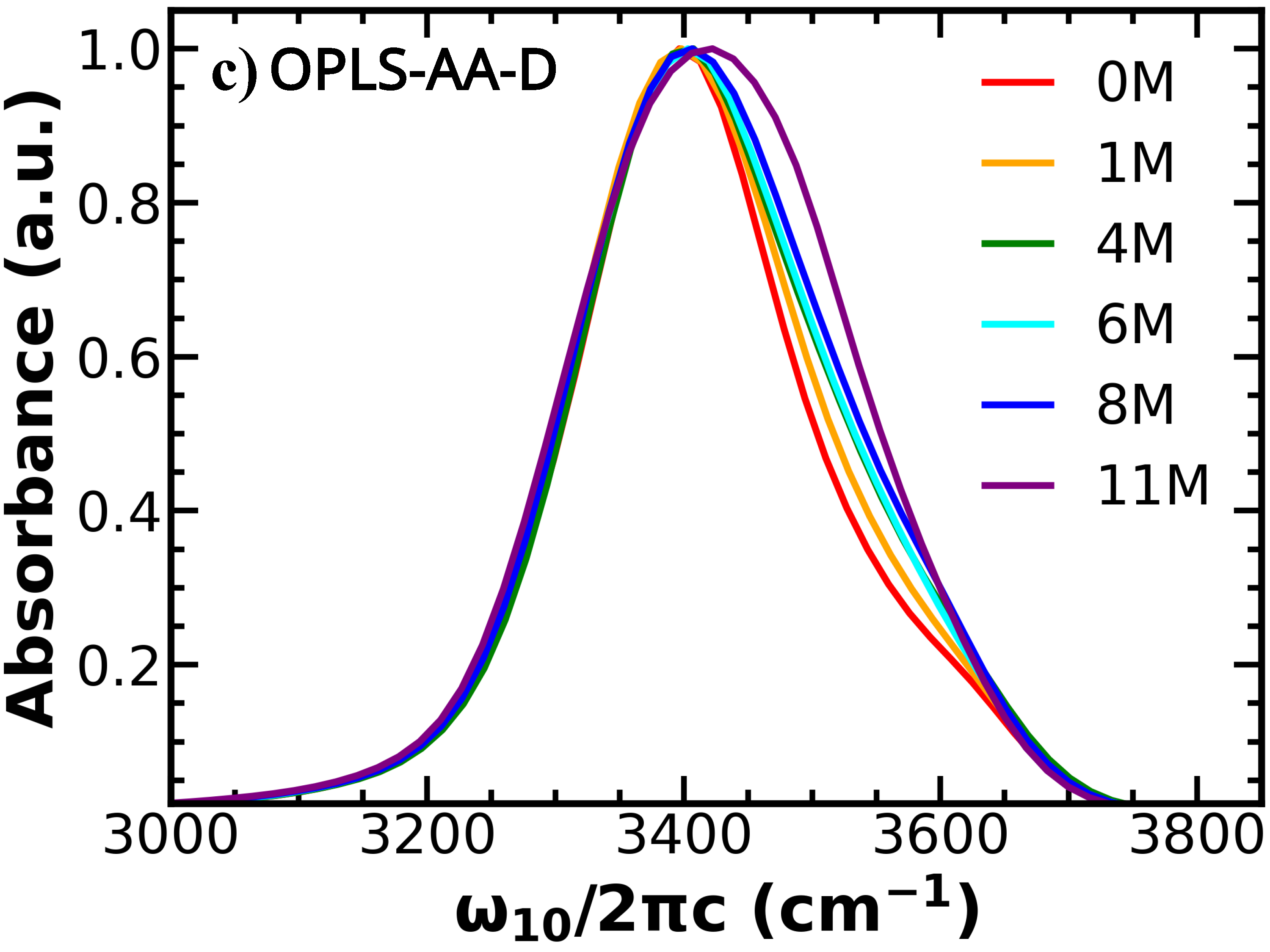}
    \includegraphics[width=0.45\textwidth, clip]{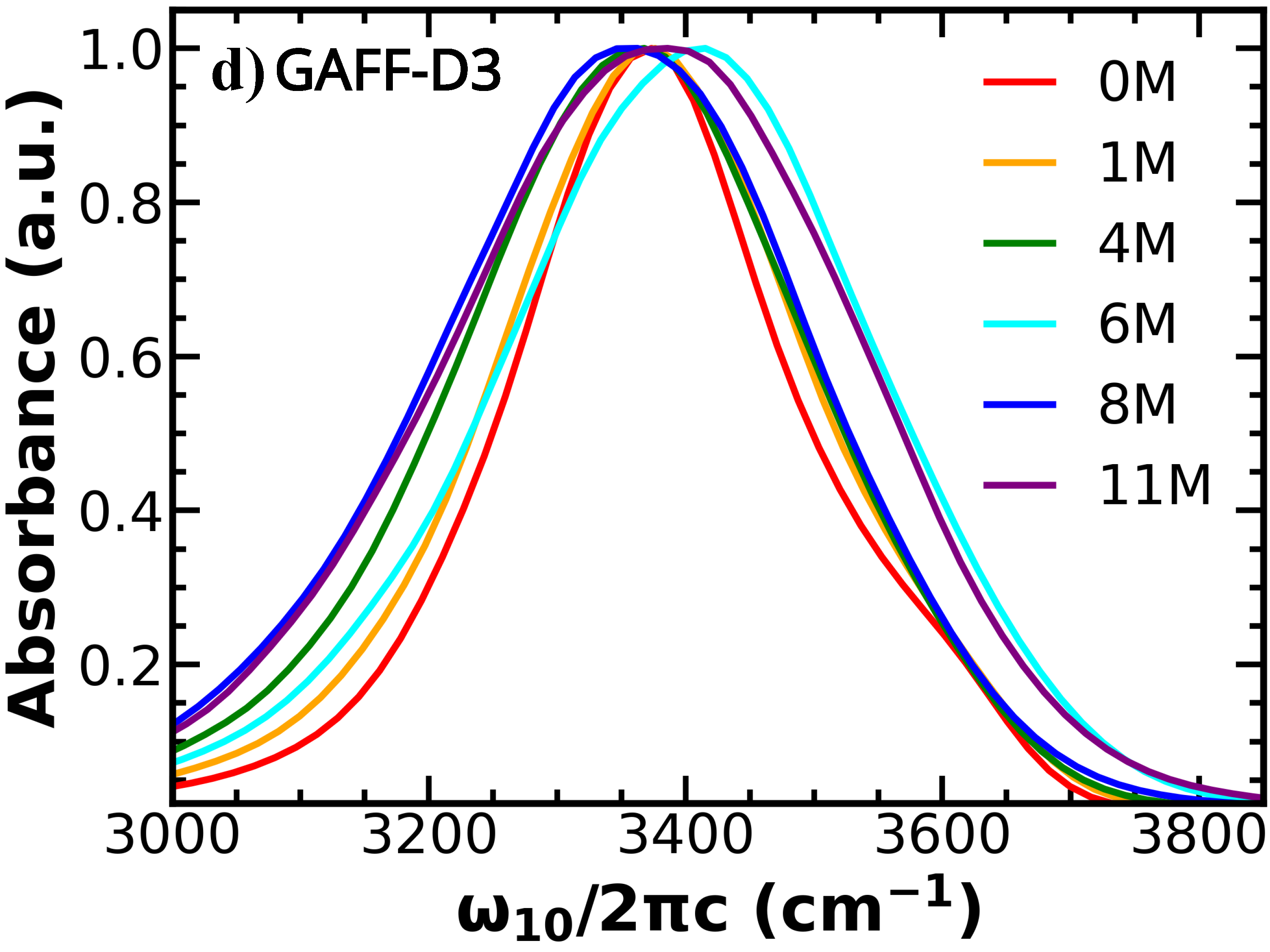}
    \caption {Comparison of 1D infrared spectra of OH stretch from various concentrations of urea between different force field of urea, \textbf{a)} KBFF,  \textbf{b)} OPLS-S, \textbf{c)} OPLS-AA-D, \textbf{d)} GAFF-D3.}
    \label{figure:IR}
\end{figure*}
\section{\label{sec:level3}Results and Discussions}
\subsection{\label{sec:level31}Linear IR Spectra of the Water OH Stretch}
The linear infrared (IR) spectrum of the isolated OH stretch is a highly sensitive probe of the local environment, and thus an effective indicator of solute-induced structural changes in aqueous systems. External solutes generally perturb the local electronic environment of nearby solvent molecules, which is reflected in their vibrational spectra. Surprisingly, previous studies using both experimental and theoretical spectra revealed an invariance in the OD and OH stretch peak position across a range of concentrations of urea.\cite{grdadolnik2002urea,idrissi2005molecular,rezus2006effect,carr2013structure,bandyopadhyay2014molecular,HAMMAMI2021113218} In our previous study,\cite{ADHIKARY2024100609} we explained the origin of this apparent spectral inertness by microscopic spectral decomposition. Water molecules in proximity to urea’s carbonyl (C=O) group exhibited red-shifted, narrower frequency distributions indicative of stronger hydrogen bonding. In contrast, those near the amine (NH$_{2}$) group showed the opposite behaviour,blue-shifted and broader profiles. However, the net contribution of these perturbed populations to the overall IR spectrum was minimal, owing to their low abundance and the counteracting nature of their spectral shifts. In this section, we explore the IR spectral response of the OH stretch in urea-water solution for different force fields of urea ( KBFF,  OPLS-S, OPLS-AA-D, GAFF-D3 ). In the absence of lifetime effects, the one-dimensional absorption lineshape can be obtained by taking the Fourier transform of the real part of the linear response function  $R^{(1)}(\tau)$, \cite{Auer2008, SCHMIDT2007143,Lin2009NaBr,mukamel1995principles} as given below. 
\begin{equation}
    R^{(1)}(\tau) = \left\langle \mu_{nm}(\tau) || \mu_{nm}(0) \exp\left[-i\int_{0}^{\tau} \omega_{nm}(t) \, dt\right] \right\rangle
\end{equation}
We consider the average response from multiple O-H trajectories to compute the linear IR response for bulk water.
\par

For urea-water systems, we obtain the FTIR response by tagging O-H oscillators based on their cumulative residence time (50 - 100 ps ) in the first solvation shell of urea. The calculated spectra for the bulk water and various urea-water systems are shown in Figure \ref{figure:IR}. The spectra from the KBFF force field of urea maintain the bulk water response even at very high urea concentrations (Figure \ref{figure:IR}\textbf{a}); however, the response from both versions of Duffy (OPLS-S Figure \ref{figure:IR}\textbf{b}, OPLS-AA-D Figure \ref{figure:IR}\textbf{c}) and GAFF-D3, Figure \ref{figure:IR}\textbf{d} is perturbed by the increasing concentration of urea. Spectral response from both OPLS-S and OPLS-AA-D is blue-shifted at high urea concentrations; in contrast, the trend of spectral response from GAFF-D3 is random, and clear spectral broadening is observed for high urea concentrations. However, earlier experimental and computational findings \cite{grdadolnik2002urea,rezus2006effect,ojha2019urea,carr2013structure} confirmed that linear IR spectra of O-H and O-D stretch are invariant with increasing urea concentration. Therefore, from Figure \ref{figure:IR} it is obvious that only the KBFF force field of urea matches the earlier experimental and computational outcome.\cite{rezus2006water,rezus2006effect, carr2013structure, ADHIKARY2024100609} Further analyses of the limitations observed in other urea force fields (OPLS-S, OPLS-AA-D, and GAFF-D3) prompted a detailed investigation of urea–urea and urea–water interactions.

\begin{figure}[htbp!]
    \centering
    \includegraphics[width=0.49\textwidth, clip]{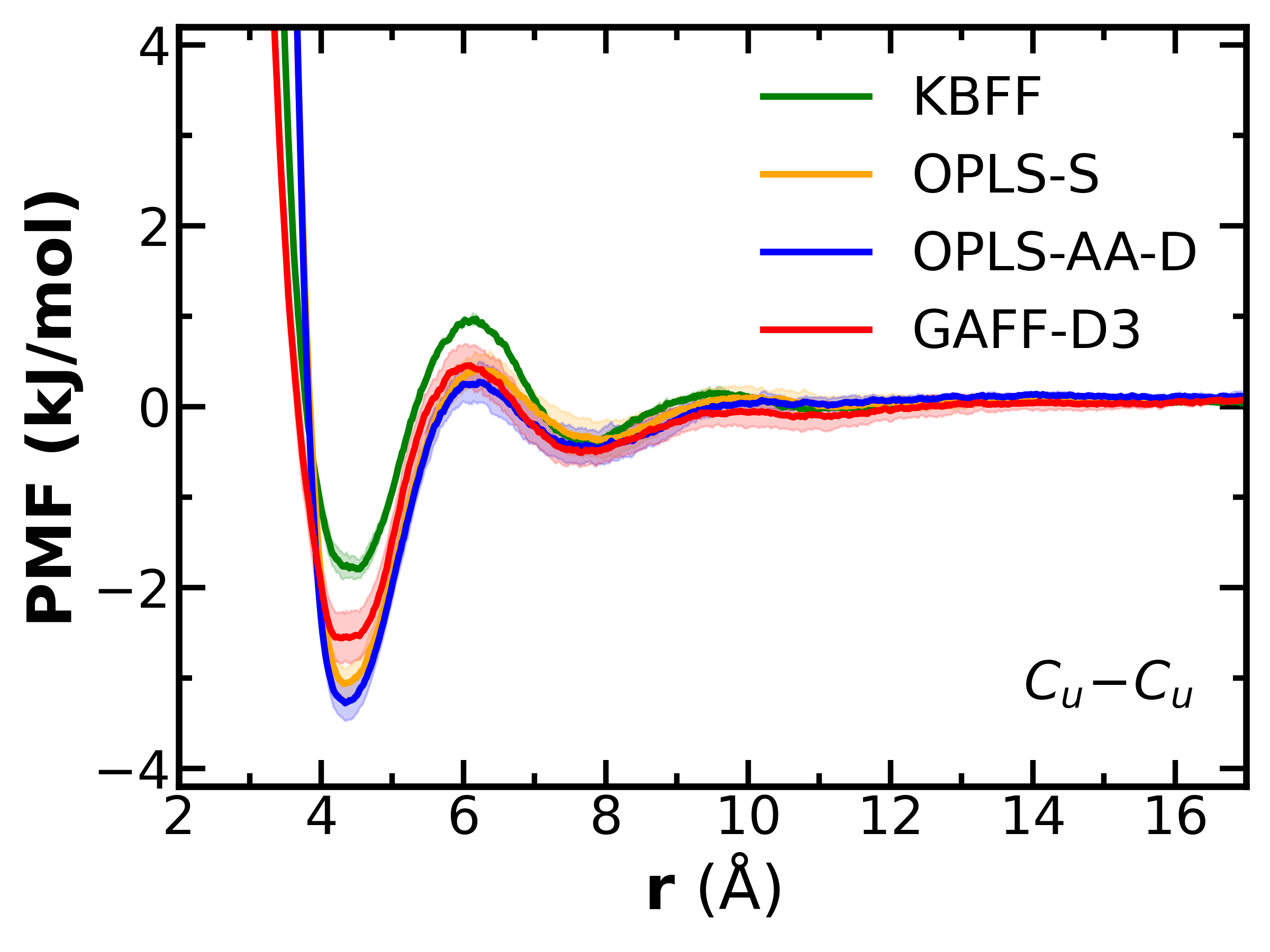} 
    \includegraphics[width=0.49\textwidth, clip]{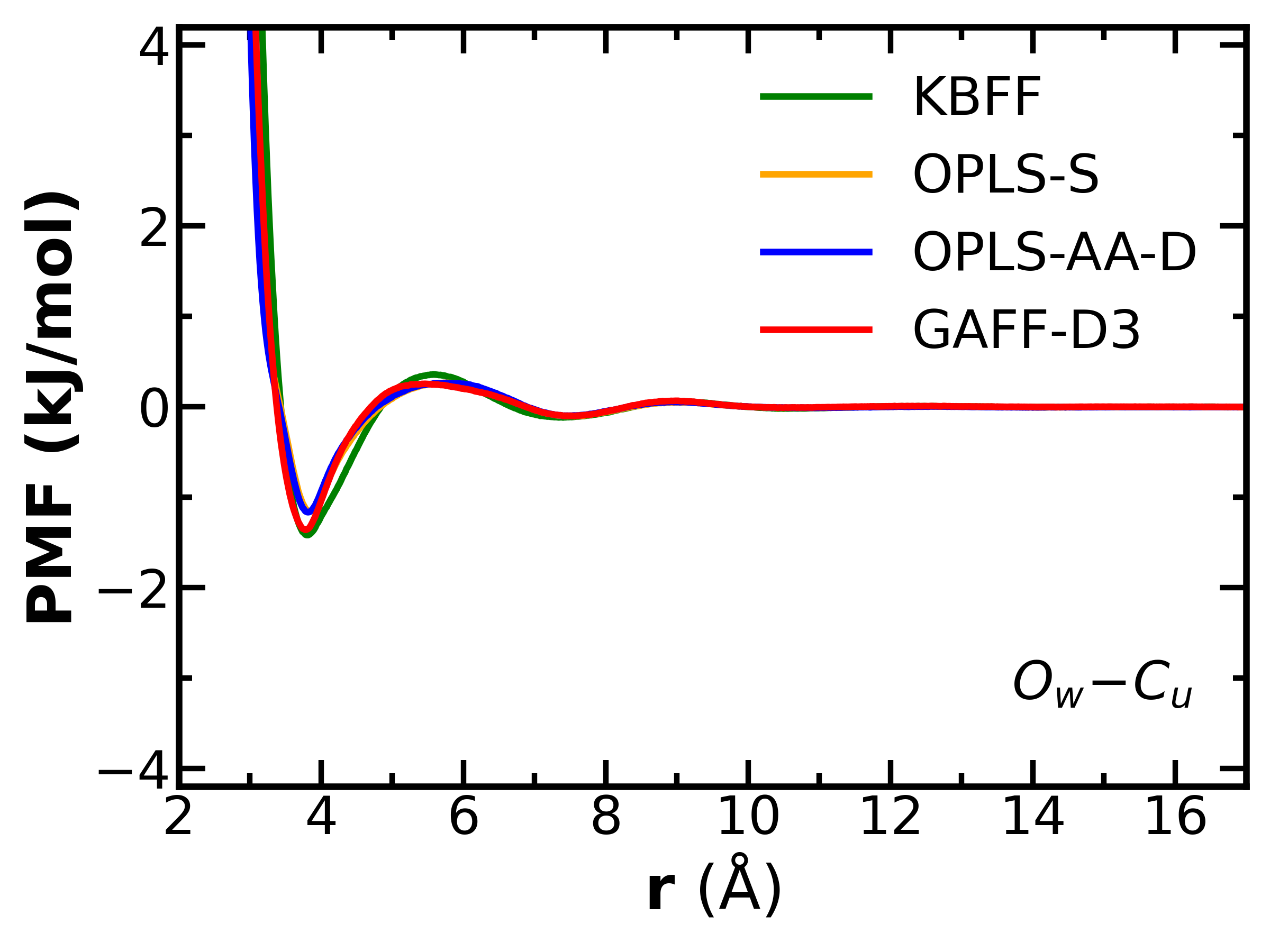}
    \caption {Potential of mean force between \textbf{a)} urea-urea, and \textbf{b)} urea-water calculated from 1M urea-water solution of various urea force fields.}
    \label{figure:pmf}
\end{figure}

\subsection{\label{sec:level32}Potential of Mean Force}
In order to understand the stability of urea-urea and urea-water interactions of various urea force fields, we calculate the potential of mean force (PMF) profiles for urea-urea and urea-water pairs using the C$_u$–C$_u$ and O$_w$–C$_u$ radial distribution functions, respectively, from the standard Boltzmann-inversion relation.\cite{bagchi2013water}
\begin{equation}
    PMF(r) = -k_BT\cdot ln(g(r))
\end{equation}

Figure \ref{figure:pmf}\textbf{a},  \ref{figure:pmf}\textbf{b} depicts the urea-urea and urea-water PMF at urea concentration 6 M, respectively. All four force fields exhibit a pronounced primary minimum around C$_u$-C$_u$ distance $\approx$ 4.40 Å as shown in the Figure \ref{figure:pmf}\textbf{a}. OPLS-AA-D shows maximum depth  (–3.01 kJ/mol) followed by OPLS-S (-2.86 kJ/mol), GAFF-D3 (- 2.06 kJ/mol) and KBFF (-1.74 kJ/mol). This indicates that urea-urea attraction is much stronger in both the OPLS force fields of urea than in GAFF-D3 and KBFF. Figure \ref{figure:pmf}\textbf{b} presents the urea-water PMF at the same concentration. All four force fields exhibit a pronounced primary minimum around O$_w$-C$_u$ distance $\approx$ 3.80 Å. KBFF shows maximum depth (–1.53 kJ/mol) followed by GAFF-D3 (- 1.39 kJ/mol), OPLS-S (-0.97 kJ/mol) and OPLS-AA-D (0.89 kJ/mol). 
\par A similar picture is observed for some other concentrations of urea. For both the OPLS models deeper C$_u$-C$_u$ minima relative to O$_w$–C$_u$ confirm that at higher urea concentration, self‐association predominates over urea–water hydrogen bonding, underpinning the clustering behavior observed at high concentrations in these models.\cite{duffy@1993, sokolic2002molecular, stumpe2007aqueous, Idrissi2008} However, KBFF and GAFF-D3 showcase a comparable urea-urea and urea-water interaction strength. To comprehensively understand this phenomenon, we evaluate and compare the interfacial surface area between urea and water in various concentrations of urea in the following section.

\begin{figure*}[htb!]
    \centering
    \includegraphics[width=0.49\textwidth, clip]{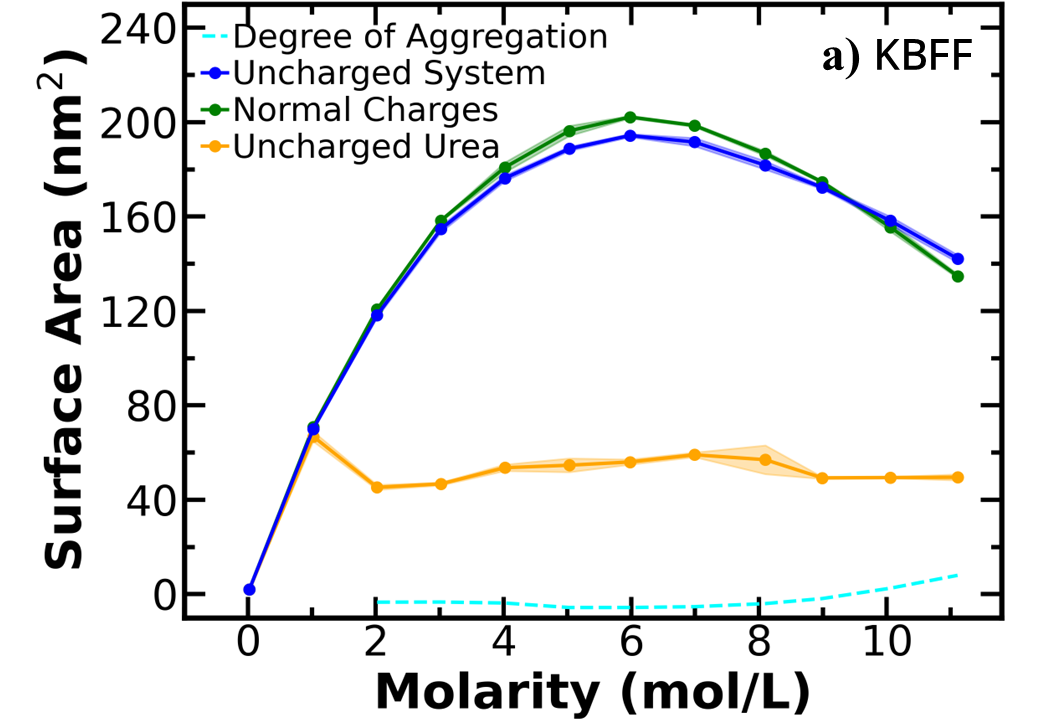} 
    \includegraphics[width=0.49\textwidth, clip]{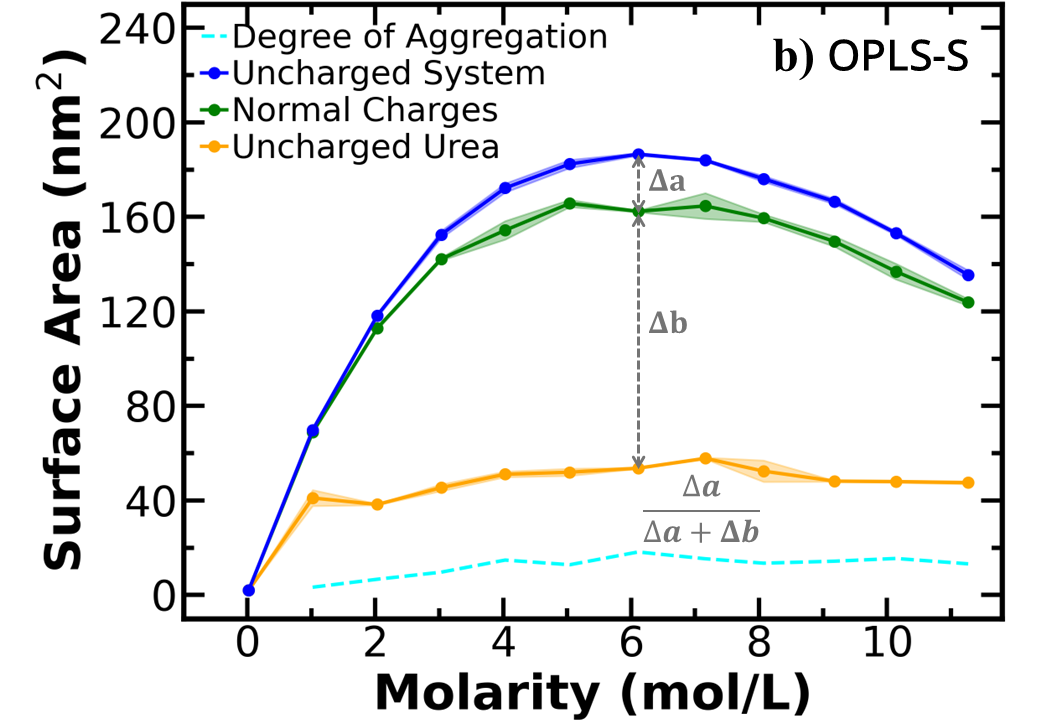}
    
        \vspace{0.25cm}  

    \includegraphics[width=0.49\textwidth, clip]{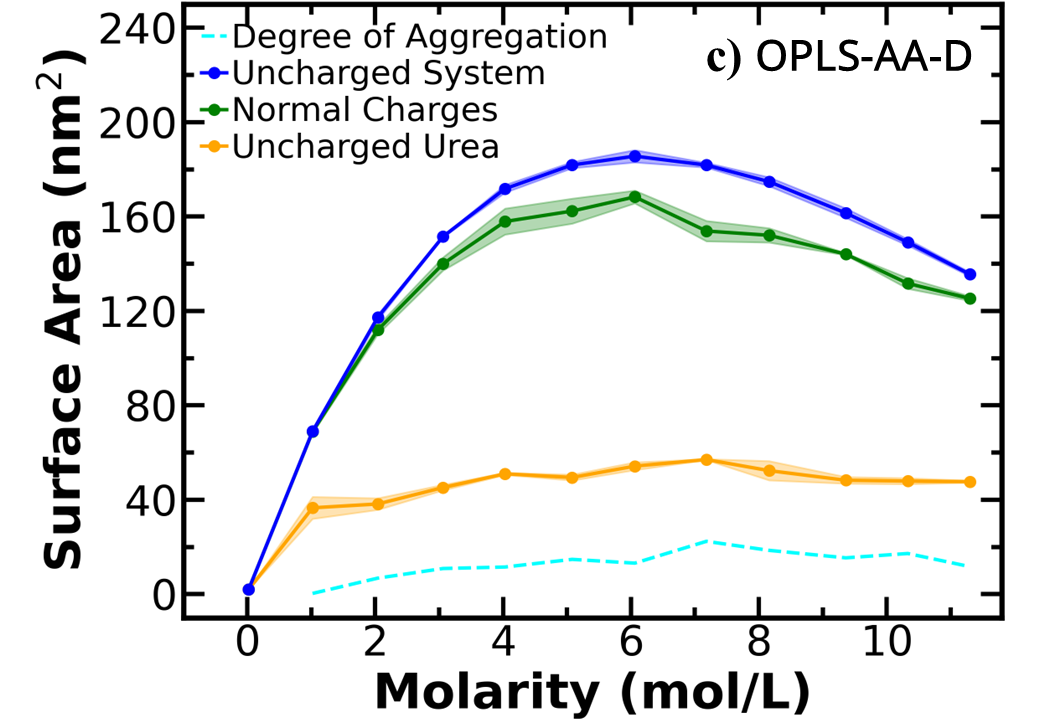}
    \includegraphics[width=0.49\textwidth, clip]{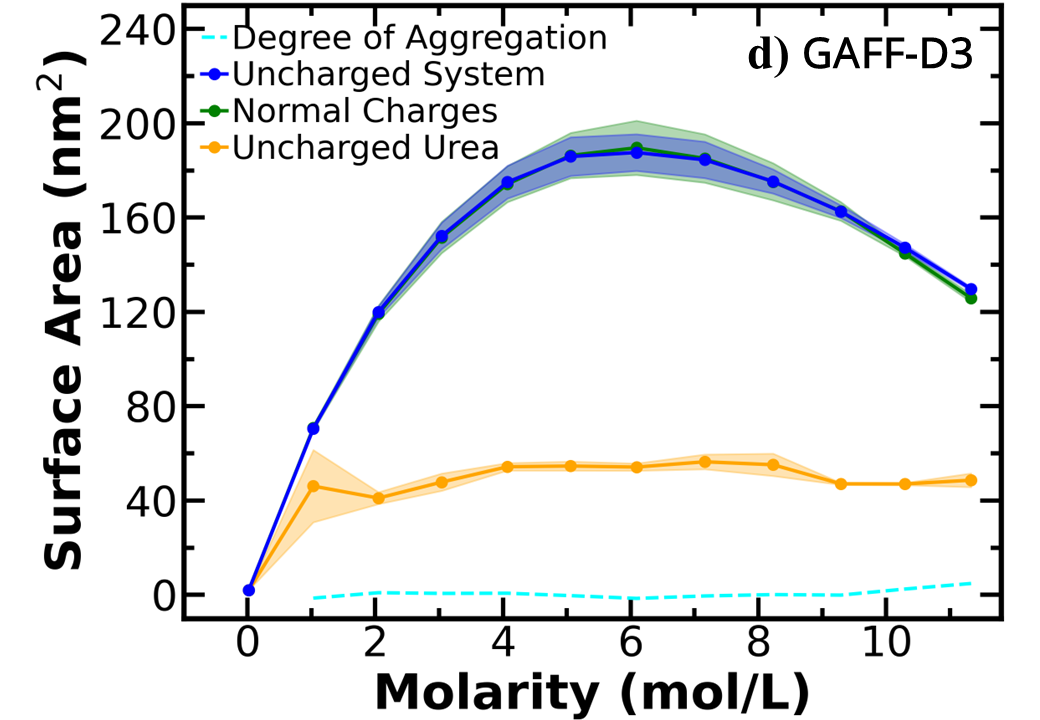}
    \caption {Comparison of Interface surface area with Normal Charges within the range of minimal (Uncharged System) to maximum aggregation (Uncharged Urea) between different force fields of urea, \textbf{a)} Kirkwood Buff Derived , \textbf{b)} OPLS-S , \textbf{c)} OPLS-AA-D , \textbf{d)} GAFF-D3.}
    \label{figure:sasa}
\end{figure*}

\subsection{\label{sec:level33}Aggregation of Urea}
The self-aggregation of urea is quantified by estimating the interface surface area between water and urea by following the technique of Stumpe et al.\cite{stumpe2007aqueous} as briefly described in the methodology section. We compare the interface surface area between urea and water of various urea-water concentrations (Normal Charges) for all four force fields of urea along with two hypothetical partial charges (Uncharged System, Uncharged Urea) referring to the lower (0\%) and upper (100\%) limit of aggregation levels as depicted in Figure \ref{figure:sasa}, by blue and orange lines with circles, consecutively. For statistical reliability and reproducibility, each observation is made three times and shown as a shaded area (standard deviation) with a solid line (mean) in Figure \ref{figure:sasa}. The interface surface area is expected to vary linearly without any random contacts between urea molecules. However, because of the limited system volume and the stochastic clustering of urea molecules, the interface surface area grows sub-linearly with increasing concentration. The decline in interface surface area above 7 mol/L urea concentration is an expected outcome, as a pure urea system would inherently lack any interfacial surface. The upper limit of aggregation (100\%) is defined using a set of simulations with completely uncharged urea while water molecules retained their standard charges (orange line with shaded area and circle in Figure \ref{figure:sasa}). The degree of aggregation within purely random clustering and maximum aggregation is illustrated by the lower (cyan dashed) line in Figure \ref{figure:sasa}. Figure \ref{figure:sasa}\textbf{a} illustrates the KBFF-SPC/E urea-water pair with standard charges (blue line with shaded area and circle), exhibiting slight repulsion between urea molecules.
In contrast, Figure \ref{figure:sasa}\textbf{b} and \textbf{c} representing OPLS-S - SPC/E and OPLS-AA-D - SPC/E pairs consecutively, show aggregation, confirming their inherent self-association behaviour, as widely observed in previous literature.\cite{smith2004computer, stumpe2007aqueous,Idrissi2008} The GAFF-D3 - SPC/E pair carries a neutral signature, distinct from the others. However, the deviation at typical concentrations used for protein denaturation (about 6-8 M), KBFF is slightly repulsive in nature with a negative degree of aggregation. In contrast, two models of duffy (OPLS-S, OPLS-AA-D) are attractive, having a positive degree of aggregation (with a maximum of 18\%, 25\%, respectively), and GAFF-D3 is neutral, having a degree of aggregation very close to zero with a high standard deviation. These observations are consistent with the PMF outcomes of the previous section \ref{sec:level32}.

\subsection{\label{sec:level34}Hydrogen Bond Strength and Population}
\begin{figure*}[htbp!]
    \centering
    \includegraphics[width=0.49\textwidth, clip]{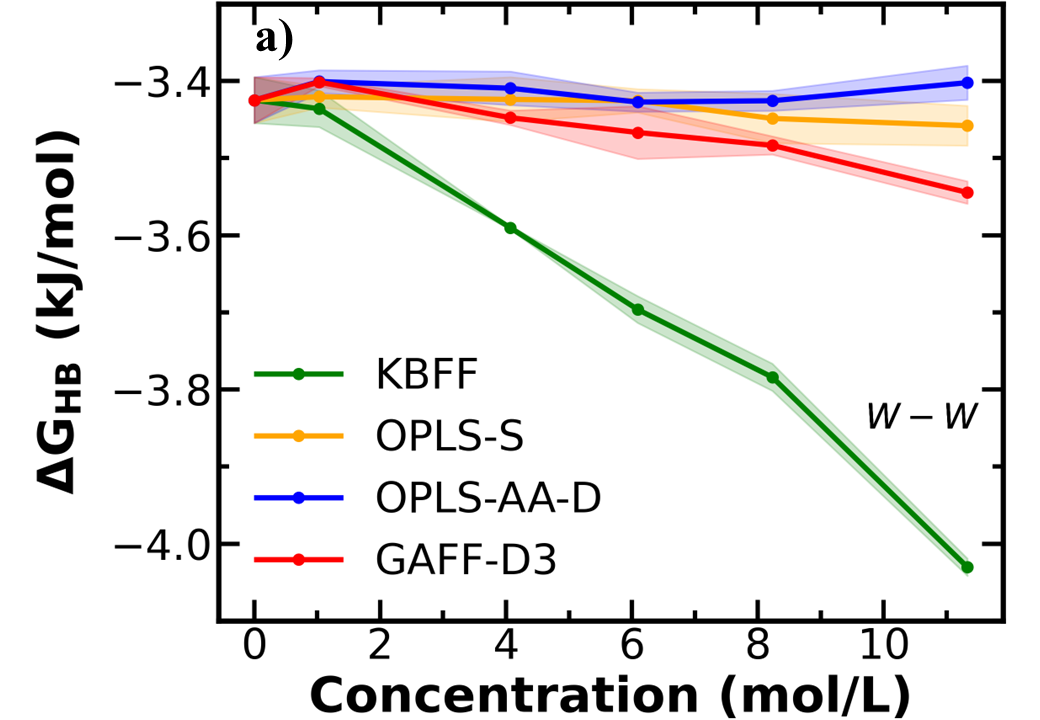} 
    \includegraphics[width=0.49\textwidth, clip]{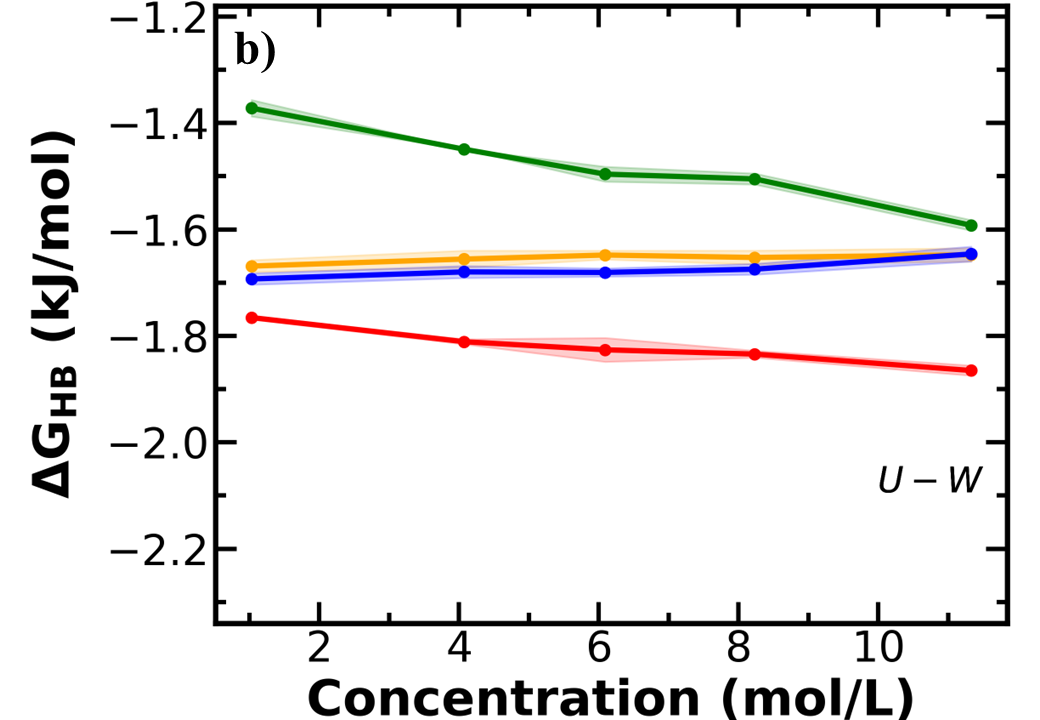}
        \vspace{0.25cm}  

    \includegraphics[width=0.49\textwidth, clip]{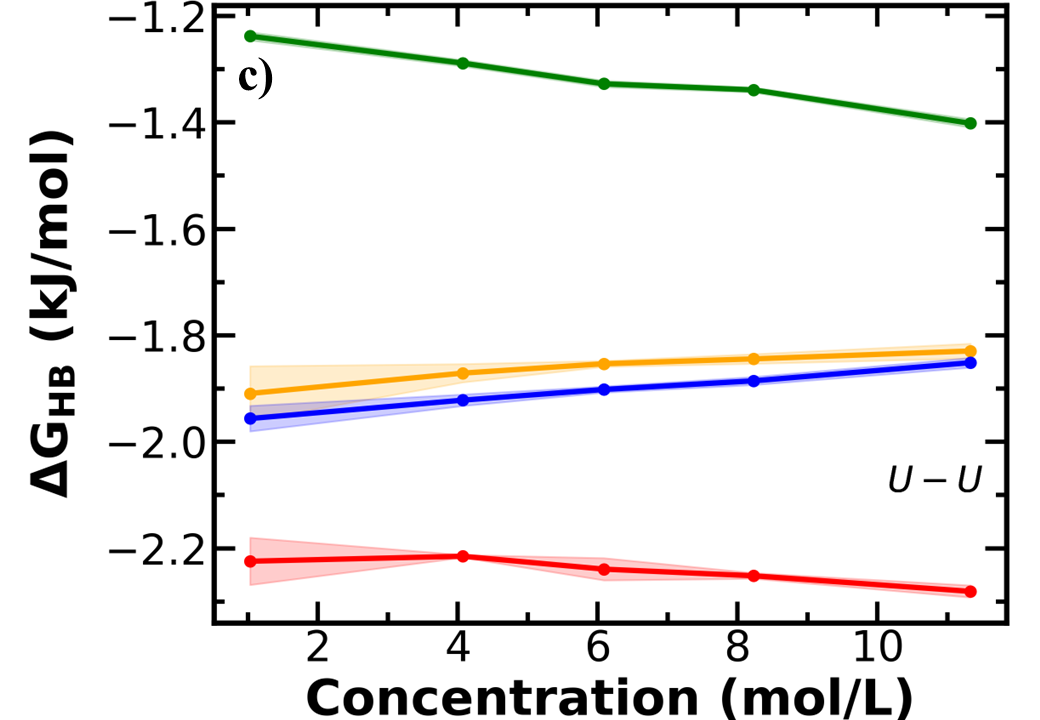}
    \caption {Variation of Hydrogen Bond Strength $\Delta G_{HB}\ $ (kJ/mol) of various interaction pairs \textbf{a)} W-W,  \textbf{b)} U-W, \textbf{c)} U-U as a function of urea concentrations between different force field of urea.}
    \label{figure:hbe}
\end{figure*}

To quantitatively assess urea’s influence on water’s H-bond network and its connection to water structure and urea aggregation, we evaluate the H-bond strength of relevant donor (D) - acceptor (A) pairs across pure water and urea solutions. We follow the methodology briefly described in section \ref{sec:level24}. We find that the W-W H-bond strength ($\Delta G_{HB}$) is more stable than U-W and U-U for all the urea models and across all the concentrations of urea as shown in Figure \ref{figure:hbe}.
This finding reinforces that urea is a weak perturbant of water's hydrogen-bonding network. \cite{rezus2006effect,carr2013structure,bandyopadhyay2014molecular,ojha2019urea,ADHIKARY2024100609} A gradual strengthening of W-W H-bond is observed with increasing urea concentration for the KBFF model, which is less prominent in GAFF-D3 and missing for OPLS-S and OPLS-AA-D models. The KBFF model has the least stable U-W, U-U H-bond, followed by OPLS-S, OPLS-AA-D and GAFF-D3 models, respectively. The Weaker of U-W H-bond than W-W in  OPLS-S, OPLS-AA-D and GAFF-D3 supports the spectral blue-shift as observed in Figure \ref{figure:IR}\textbf{a}, \textbf{b}, \textbf{c} respectively. The KBFF model reveals stronger U–W H-bonds than U–U, in contrast to the other three models where U–U H-bonds dominate, as shown in Figure \ref{figure:hbe}\textbf{b} and \textbf{c} and matches the outcome of section \ref{sec:level32}. The weakest U-U H-bond strength of KBFF supports its repulsive nature observed in Figure \ref{figure:sasa}\textbf{a}. The higher stability of the U-U $\Delta G_{HB}$ in both versions of the Duffy model supports their aggregating nature\cite{stumpe2007aqueous} observed in Figure \ref{figure:sasa}\textbf{a} and \textbf{b} sequentially. In the case of KBFF, we can see a prominent gradual increase in the stability of all three kinds of H-bond strength upon increasing concentration of urea, which corroborates its inertness towards water local structure. We've provided much more microscopic insight regarding this in our earlier work.\cite{ADHIKARY2024100609} GAFF-D3 shows a less prominent, almost similar picture with the only exception of $\Delta G_{HB}$ U-U > $\Delta G_{HB}$ U-W in terms of stability. Meanwhile, the other two models (OPLS-S and OPLS-AA-D) reveal that all three types of H-bond are almost constant when urea concentration is increased. \par
\begin{figure*}[htbp!]
    \centering
    \includegraphics[width=0.49\textwidth, clip]{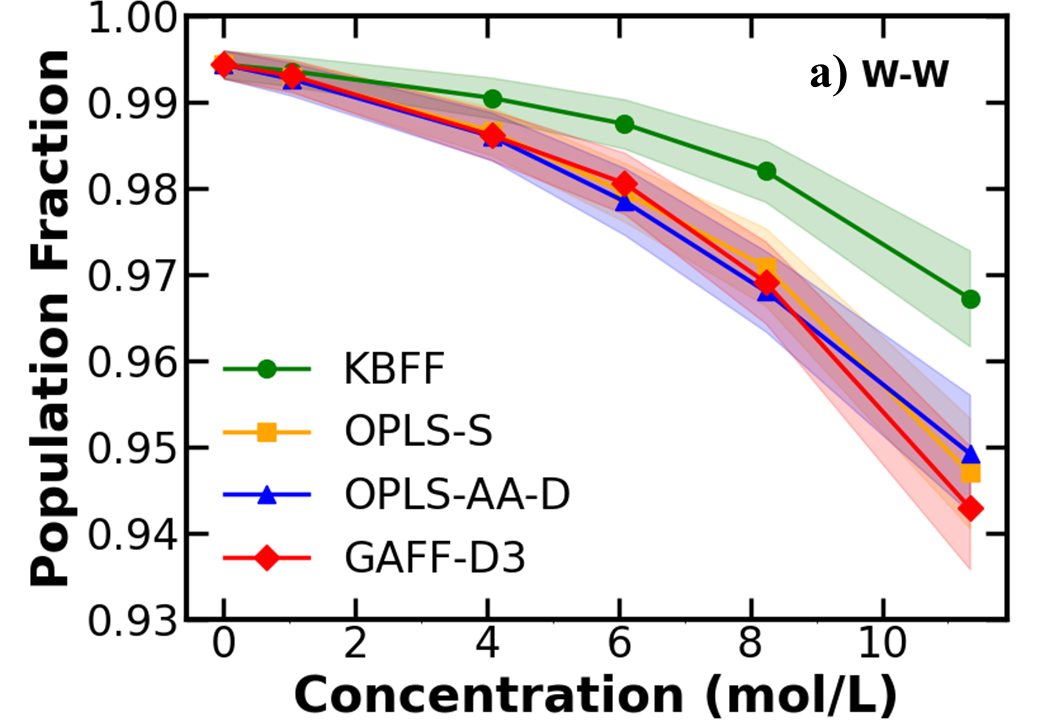}
    \includegraphics[width=0.49\textwidth, clip]{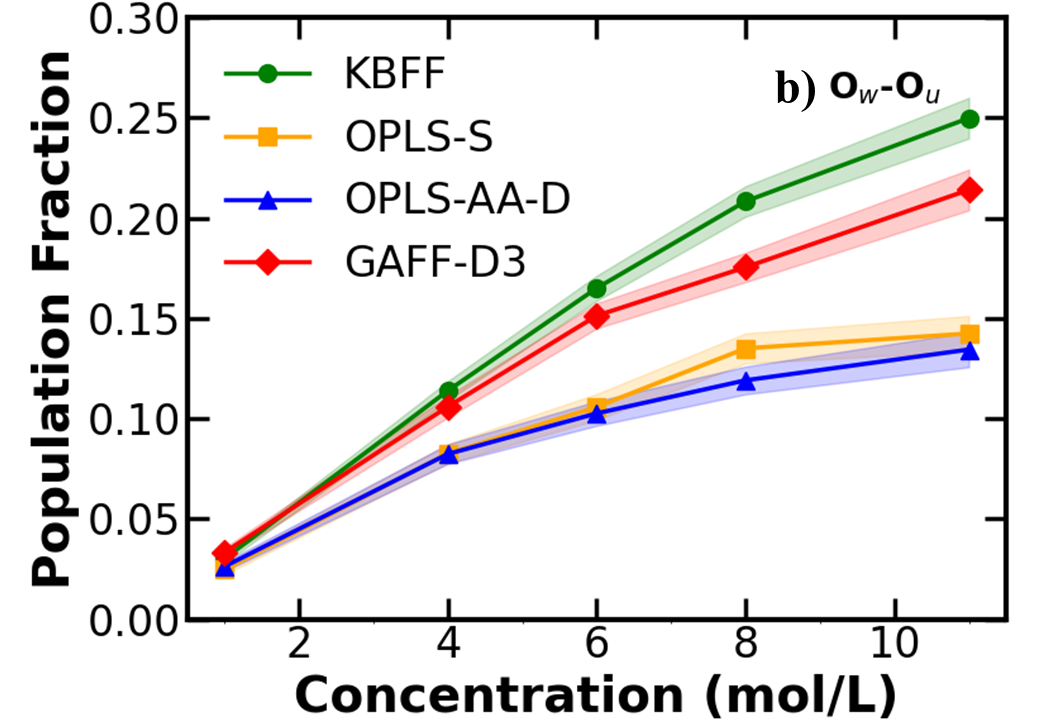}  
            \vspace{0.25cm}  
    \includegraphics[width=0.49\textwidth, clip]{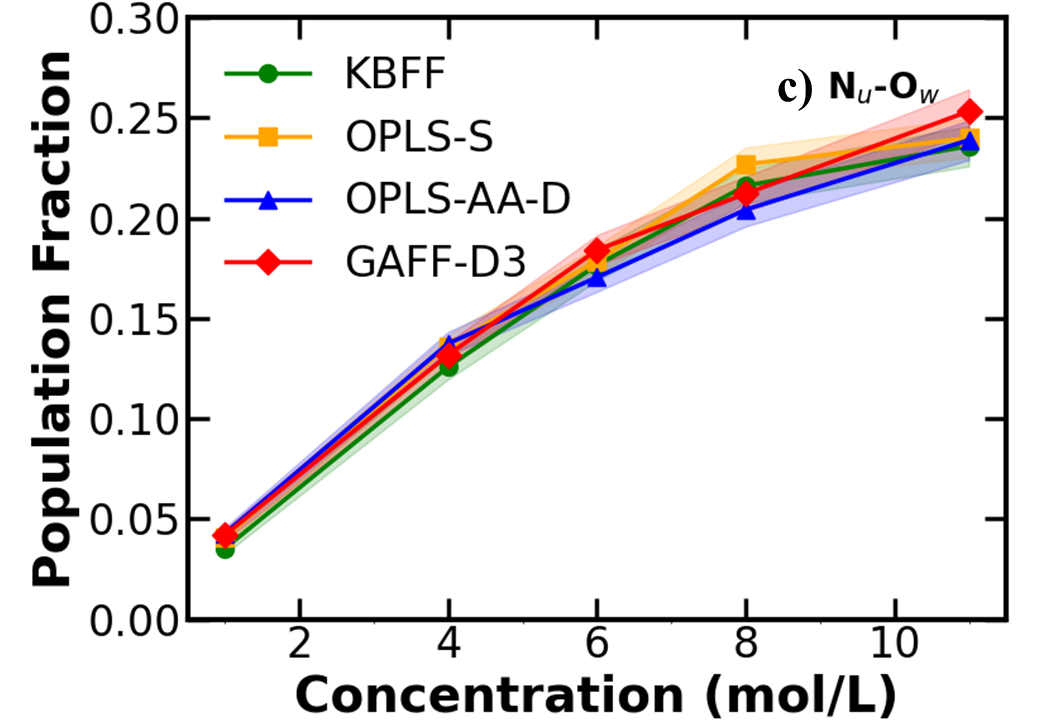}
    \caption {Hydrogen Bond population per water molecule of interaction pairs \textbf{a)} W-W (O$_w$-O$_w$), \textbf{b)} O$_w$-O$_u$,  \textbf{c)} N$_u$-O$_w$ as a function of urea concentrations between different force field of urea.}
    \label{figure:hbpop}
\end{figure*}
\begin{figure*}[htbp!]
    \centering
    \includegraphics[width=0.49\textwidth, clip]{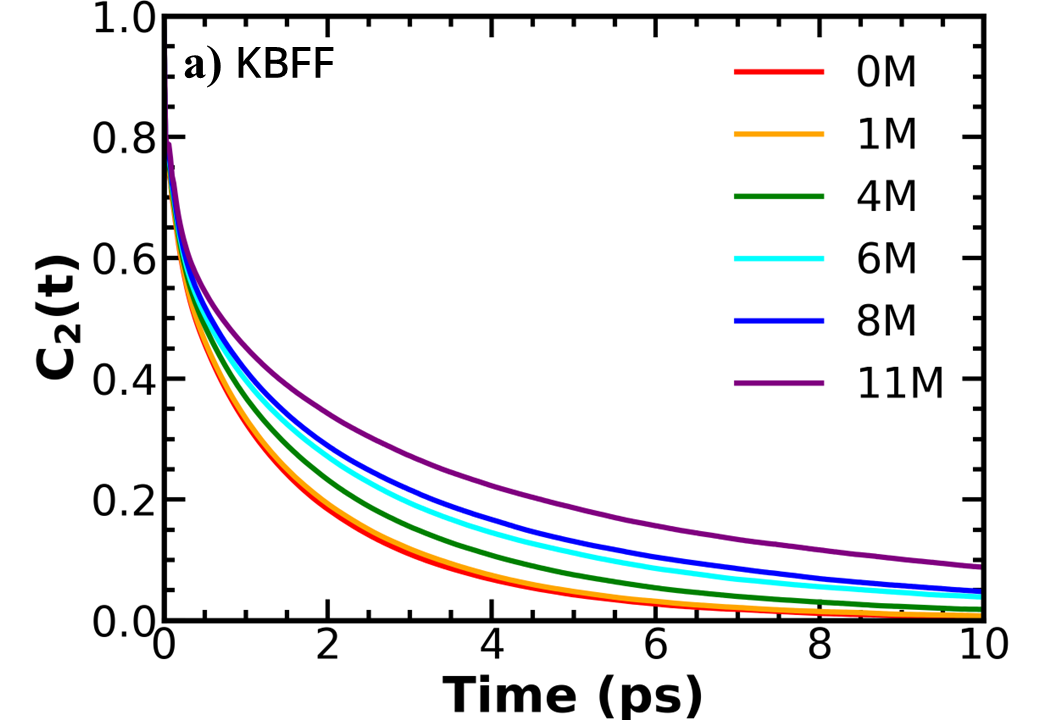} 
    \includegraphics[width=0.49\textwidth, clip]{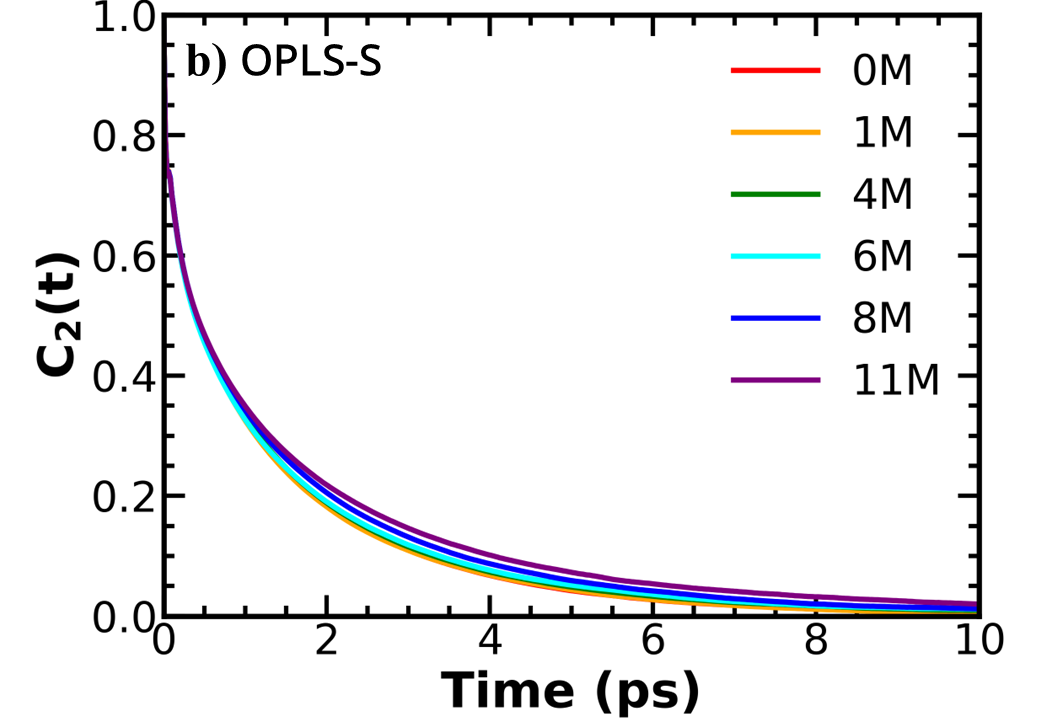}
    
        \vspace{0.25cm}  

    \includegraphics[width=0.49\textwidth, clip]{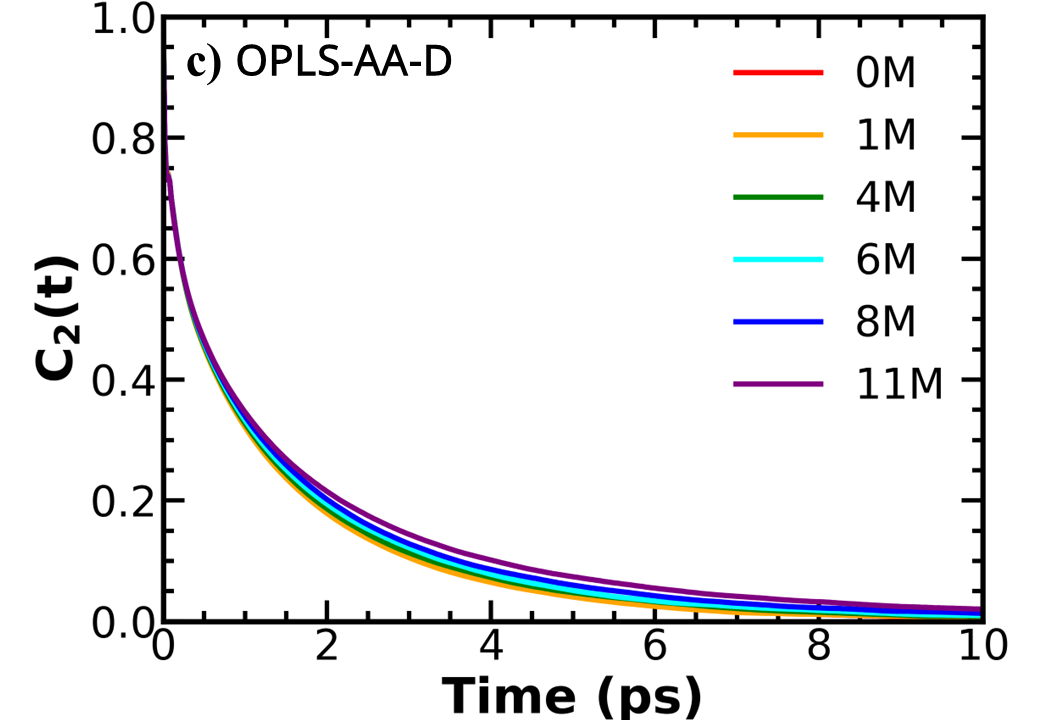}
    \includegraphics[width=0.49\textwidth, clip]{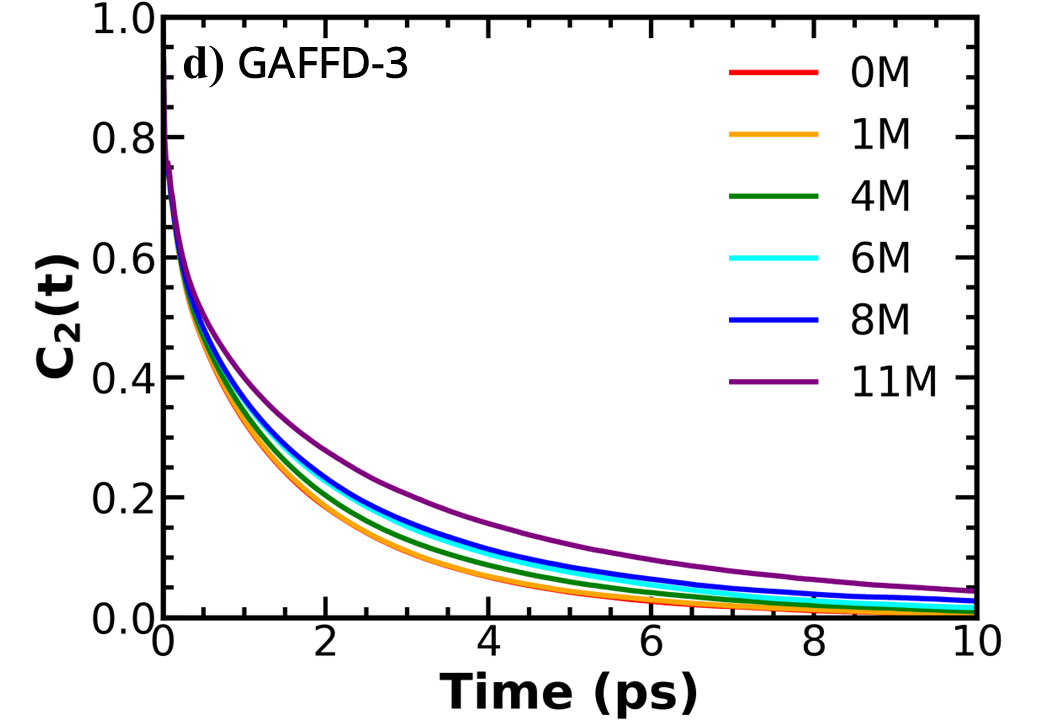}
    \caption {Orientational Time Correlation function (C$_2$(t)) of OH bond of pure water and various concentrations (1 M, 4 M, 6 M, 8 M, 11 M) of urea-water binary solution across different force fields of urea with fitting\textbf{a)} Kirkwood Buff Derived , \textbf{b)} OPLS-S , \textbf{c)} OPLS-AA-D , \textbf{d)} GAFF-D3.}
    \label{figure:otcf}
\end{figure*}
We further analyzed W–W (O$_w$–O$_w$) H-bond population in pure water and urea–water mixtures of various concentrations. Additionally, we examined U–W H-bond populations in detail by evaluating the O$_w$–O$_u$ and N$_u$–O$_w$ interactions in various urea-water solutions across all urea models, as shown in Figure \ref{figure:hbpop}. The population fraction is obtained by first calculating the frame-averaged total number of H-bonds in a similar manner as described in section \ref{sec:level24} for each type and finally dividing it by the total number of water molecules present in the system. It is clear from the Figure \ref{figure:hbpop}\textbf{c} that the N$_u$-O$_w$ population fraction is almost similar for all the urea models across all urea concentrations. Figure \ref{figure:hbpop}\textbf{b} reveals that the KBFF model of urea has the highest O$_w$-O$_u$ H-bond population, followed by the GAFF-D3 model, and both versions of Duffy show minimum population. However, inspecting Figure \ref{figure:hbpop}\textbf{a}, we find that the W-W H-bond population is least perturbed by the KBFF model even at very high concentration of urea, followed by the other three models of urea. This observation corroborates the IR spectral response in Figure \ref{figure:IR} \textbf{a}. 

\subsection{\label{sec:level35}Orientational Dynamics of Water}
Previous studies\cite{rezus2006effect,carr2013structure,agieienko2016urea} showed that urea induces a subtle yet noticeable slowdown in water's rotational dynamics. Here we look into water's reorientational dynamics in the presence of urea at various concentrations to examine the sensitivity of different force fields of urea. We calculate the ensemble averaged second-order orientational correlation function (C$_2$(t)) of O-H bond vectors of water molecules for all four force fields of urea as plotted in Figure \ref{figure:otcf}.

\begin{figure}[htbp!]
    \centering
    \includegraphics[width=0.49\textwidth, trim= 1.5cm 1cm 2.5cm 2.5cm, clip]{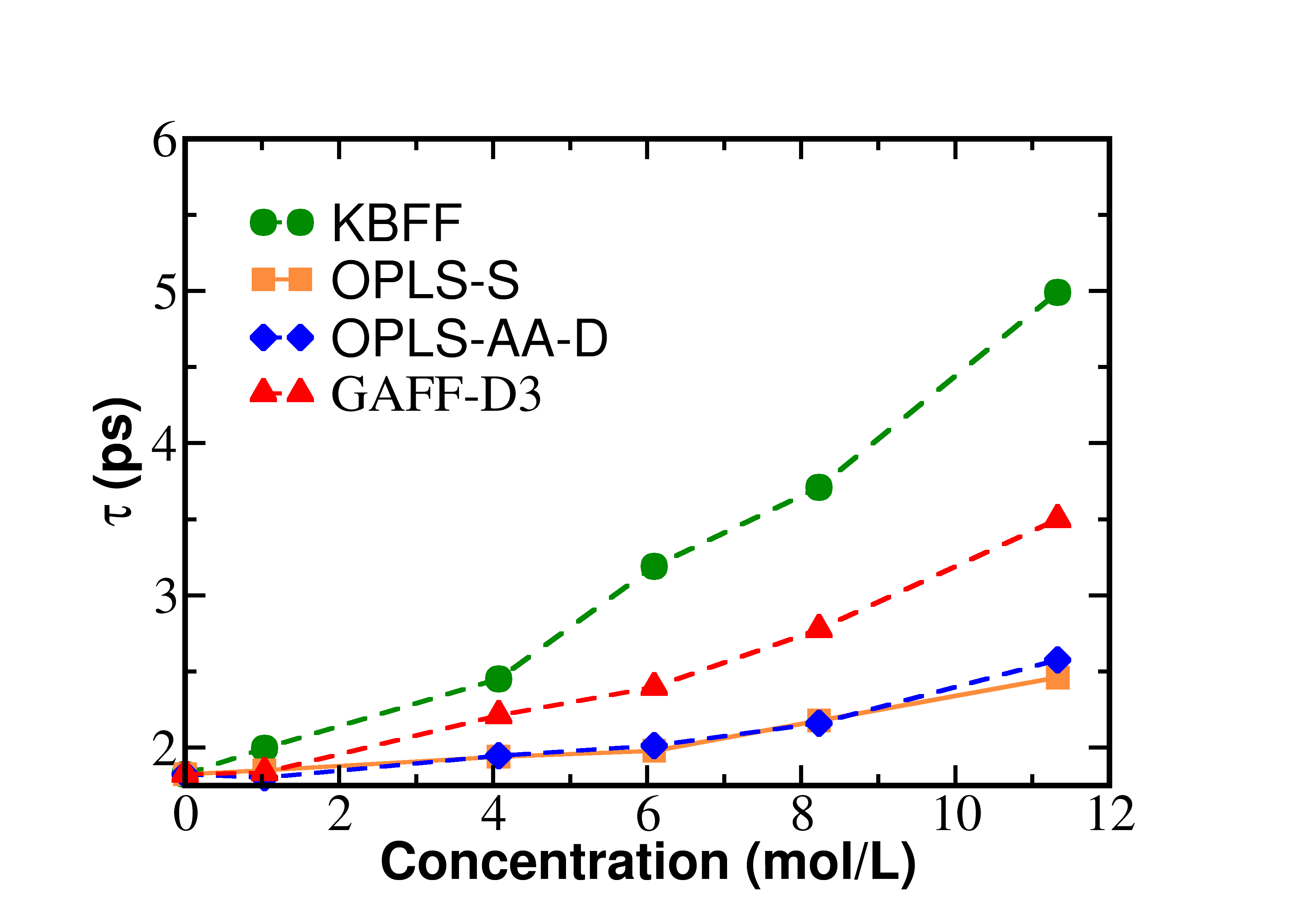} 
    \caption {Orientational relaxation time constant $\tau$ in ps as a function of urea concentration derived from bi-exponential fitting of C$_{2}$(t).}
    \label{figure:tau}
\end{figure}

To obtain the re-orientational relaxation time constant of water in both pure water and binary aqueous solutions, we fit each C$_2$(t) plot with a bi-exponential function eq \ref{eq:biexponential} to get the accurate decay times as depicted in Figure \ref{figure:tau}.
\begin{equation}
y = y_0 + a_1 e^{-x / \tau_1} + a_2 e^{-x / \tau}
\label{eq:biexponential}
\end{equation}
The relaxation time constant for bulk water is 1.82 ps, which is very close to the experimental observation of 1.7 ps obtained from rotational anisotropy measurement.\cite{rezus2006water}
 The C$_2$(t) plot, along with the fitting curve, is shown in the supporting information (FIG \textbf{S6}). We use another popular fitting function, a mixture of bi-exponential and damp oscillation exponential.\cite{moller2004hydrogen} However, we find a similar trend as bi-exponential fitting with a bit of elevated relaxation time and reduced relaxation coefficients. Water rotational relaxation time in all four models of urea show a monotonous increasing trend as a function of urea concentration as presented in Figure \ref{figure:tau}; however, a pronounced increase is observed in the relaxation time in KBFF--SPC/E (Figure \ref{figure:otcf}\textbf{a}) binary solution followed by GAFF-D3--SPC/E (Figure \ref{figure:otcf}\textbf{b}) solution. However, OPLS-S Figure \ref{figure:otcf}\textbf{c} and OPLS-AA-D--SPC/E Figure \ref{figure:otcf}\textbf{d} do not show such a signature; rather, the decay time is almost invariant with increasing urea concentration. Some recent and previous studies suggested that this distinct behaviour in water rotational dynamics can be elucidated in terms of micro-heterogeneity induced by molecular aggregation in these two urea models.\cite{rezus2006effect,Rezus2007,Hishida2022,Seo2025} The aggregation of urea in these two models leads to a nonuniform distribution of osmolytes and microscopic phase separation, resulting in confined water regions that exhibit bulk water-like rotational dynamics.

\section{\label{sec:level4}Conclusions}
External solutes typically disrupt the hydrogen-bond network, a modification sensitively captured by vibrational spectroscopy. However urea is almost inert towards water structure even at higher concentrations. In this work, we compare four widely used models of urea based on their self-association behavior and its correlation with preserving water structure and slowing down water’s rotational dynamics. Infrared spectral response reveals that three models of urea (OPLS-S, OPLS-AA-D, GAFF-D3) are not able to reproduce experimental outcomes at even moderate urea concentrations ($< 4$ M) and fail miserably at higher urea concentrations. By close inspection, through the lens of the potential of mean force, followed by the interface surface area calculation, we find that two versions of Duffy (OPLS-S, OPLS-AA-D) show the strongest C$_u$-C$_u$ interaction, resulting in self-aggregation of urea with a maximum degree of aggregation of 18\% and 25\%, respectively. Other researchers have previously reported this overestimation of the urea aggregation in these models.\cite{sokolic2002molecular,weerasinghe2003kirkwood,stumpe2007aqueous} Whereas the GAFF-D3 model reacts as neutral, and the KBFF acts as a minor repulsive in terms of the interface surface area. H-bond strength of W-W, U-W, and U-U, and the population analysis provides a more quantitative picture of self-aggregation and how it is connected to IR spectral shift and broadening. The average W–W H-bond population fraction is more perturbed in the OPLS-S, OPLS-AA-D, and GAFF-D3 models than in KBFF. This likely results from weaker U–W H-bonds compared to U–U H-bonds in these models, which supplements the spectral blue-shift and urea–urea self-aggregation. 
\par 
The deceleration of the water molecules re-orientational dynamics is captured well by the KBFF model followed by GAFF-D3, whereas OPLS-S and OPLS-AA-D doesn't show any prominent retardation. The orientational dynamics of water molecules in the presence of these two urea models is not perturbed, and almost bulk-like dynamics is observed. 
\par 
We find that the KBFF model performs best in maintaining the water structure at high urea concentration and capturing the deceleration of the orientational dynamics of water. GAFF-D3 was optimized for crystal phase; some recent work\cite{anker2023assessment} suggested it performs well in solution phase as well; however, our observations reveal that structurally and dynamically, it's not an optimum choice in aqueous solution. Both versions of Duffy (OPLS-S, OPLS-AA-D) fail in both scenarios. Therefore, we can conclude that if an urea model overestimates self-aggregation, it will affect the water structure and dynamics unrealistically, and slight repulsiveness between urea molecules at denaturing concentration is the key to realistic modelling of urea.

\section*{Acknowledgments}
R.B. acknowledges IIT Tirupati for computational support. R.B. also acknowledges the Science and Engineering Research Board, Department of Science \& Technology, India, for generous support through Grant No. CRG/2021/003859. P.A. (PMRF ID-3203664) thanks the Ministry of Education, Government of India, for the Prime Minister’s Research Fellowship for funding their doctoral studies, which made this research possible. We also acknowledge Prof. Helmut Grubmüller for insightful discussion on the interface surface area scheme.

\section*{Data Availability Statement}
The data that support the findings of this study are available from the corresponding author upon reasonable request.
\bibliographystyle{aipnum4-2}
\bibliography{reference}
\vfill\null 
\balance 
\newpage 
\clearpage 
\onecolumngrid
\section*{Supplementary Material}
\setcounter{section}{0}
\setcounter{figure}{0}
\setcounter{table}{0}
\setcounter{equation}{0}
\renewcommand\thesection{S\arabic{section}}
\renewcommand\thefigure{S\arabic{figure}}
\renewcommand\thetable{S\arabic{table}}
\renewcommand\theequation{S\arabic{equation}}

\section{Cluster Selection}
The clusters are specific spherical portions of the system. We have extracted the clusters by identifying a central oxygen atom that belongs to the water ($O_w$) molecule closest to the oxygen of urea $(O_u)$ for the -CO (Carbonyl) ensemble and nitrogen of urea $(N_u)$ for the -NH (Amine) ensemble and includes any molecules having its oxygen within a 7.0 Å radius. The length and angle cutoff for the clusters are as follows, -CO ensemble: $\textit{r}_{{O_{u}}-{O_{w}}}\leq 3.2\text{\AA}$ ($1^{st}$ solvation shell minima of $N_{u}-O_{w}$ radial distribution function) and $\angle O_{u}-H_{w}-O_{w} > 150^\circ$ -NH ensemble: $\textit{r}_{{N_{u}}-{O_{w}}}\leq 4.44\text{\AA}$ ($1^{st}$ solvation shell minima of $N_{u}-O_{w}$ radial distribution function ) and $\angle H_{u}-H_{w}-O_{w} > 150^\circ$. The average number of water molecules inside the cluster is $\approx 50$. We follow a similar approach when selecting the bulk water cluster.\cite{ADHIKARY2024100609}

\begin{figure}[htbp!]
    \centering
        \includegraphics[width=0.5\textwidth, trim=0cm 0.8cm 0cm 0cm, clip]{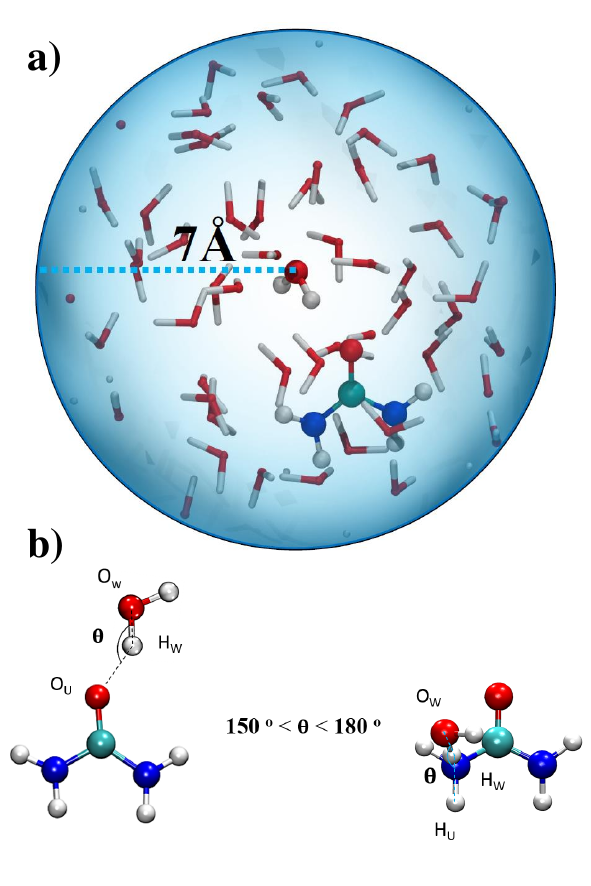}
    \caption{\textbf{a)} Urea-water cluster, \textbf{b)} The angle between water and -CO, -NH cites of urea, respectively.}
     \label{figure:cluster}
\end{figure}

\section{Collective Variable Electric Field (E)}
The electric field collective variable\cite{corcelli2005infrared} E is determined by the following expression for all the systems,
\begin{equation}
E = \mathbf{\hat{u}} \cdot \sum_{i=1}^{ab} \frac{q_{i} \hat{r}_{iH}}{r_{iH}^2}
\end{equation}
Where $\mathbf{\hat{u}}$ is the unit vector along the O-H bond of interest, $q_i$ is the charge of the ${i}$-th site, the unit vector along the direction of the distance between the ${i}$-th site and the $H$ of central $H_2O$ molecule is referred as $\hat{r}_{iH}$, the sum is taken over "$a$" molecules having "$b$" charged atoms per molecule. Here, in the case of urea, b is 4, and for water, 3.

\section{Construction of Solvation Shells}
\begin{figure}[htbp!]
    \centering
    \includegraphics[width=0.5\textwidth, trim=0.5cm 0cm 2.5cm 2cm, clip]{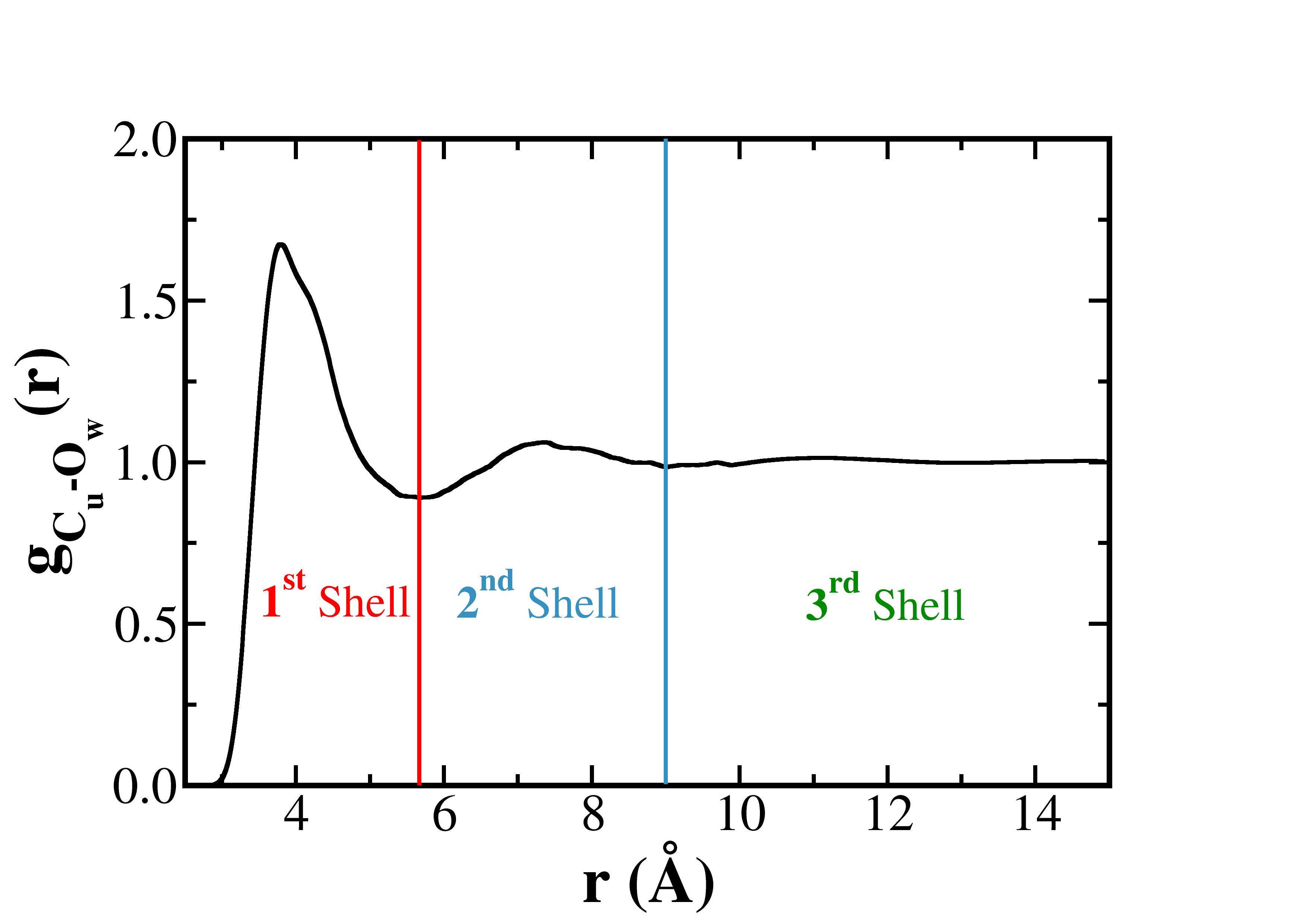}
    \includegraphics[width=0.32\textwidth, trim=0.5cm 0cm 0.5cm 0.5cm, clip]{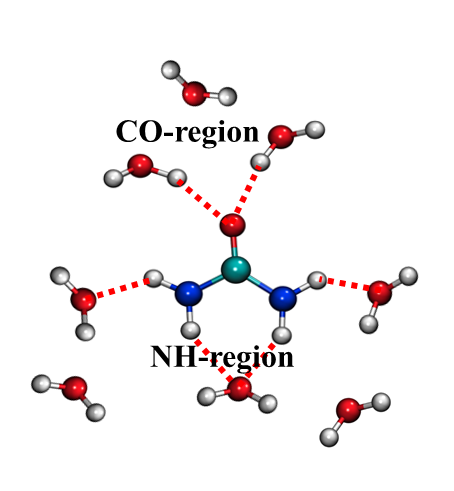}
    \caption{Pair distribution function of $C_u$ and $O_w$ for 0.21M urea system. Note that three ensemble sections of the solvation shells are highlighted in the figure.}
    \label{figure:rdfcuow}
\end{figure}

The water molecules near the urea moiety are divided into three solvation shells based on the $C_u-O_w$ pair distribution function (FIG. \ref{figure:rdfcuow}). The solvation shells are defined as $1^{st}$ shell: water molecules within 5.67  \AA{} of urea. $2^{nd}$ shell: water molecules residing within 5.67 \AA{} to 9.0 \AA{} and $3^{rd}$ shell: remaining water molecules. The $1^{st}$ shell is further subdivided into two regions of interest; the CO-region contains the OH oscillators nearest to $O_u$ (based on $O_u-O_w$ distance) for which the angle $\angle H_w-O_w-O_u$ is minimum, and the NH-region includes the OH oscillators nearest to $N_u$ (based on $N_u-O_w$ distance) for which the angle $\angle H_u-N_u-O_w$ is minimum as depicted in FIG \ref{figure:rdfcuow}.

\newpage 
\section{ Spectroscopy Maps}
\subsection{Maps for KBFF - SPC/E Pair}
\begin{figure}[H]
    \centering
    \includegraphics[width=0.45\textwidth, trim=0cm 0cm 2.5cm 1.5cm, clip]{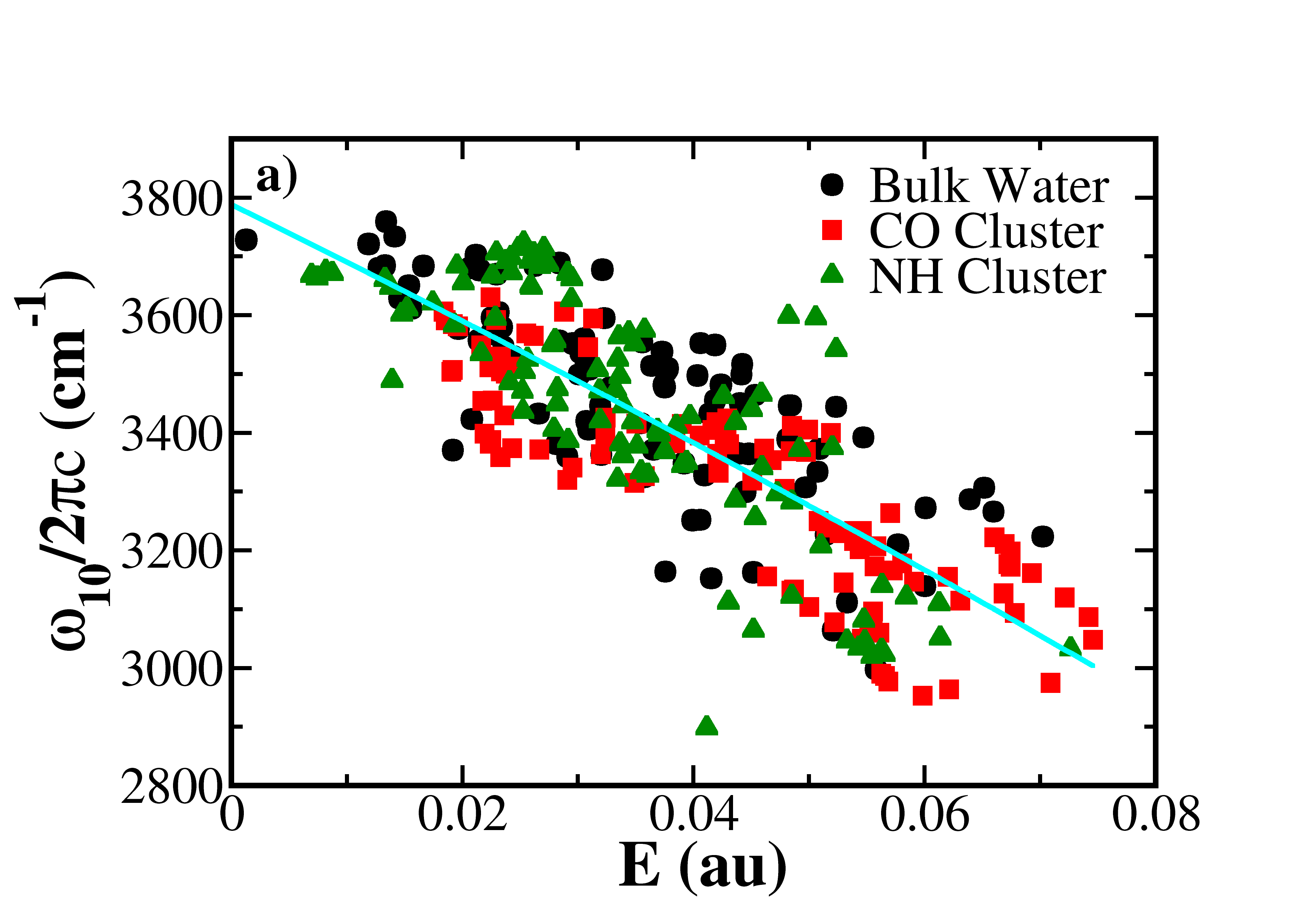} 
    \includegraphics[width=0.45\textwidth, trim=0cm 0cm 2.5cm 1.5cm, clip]{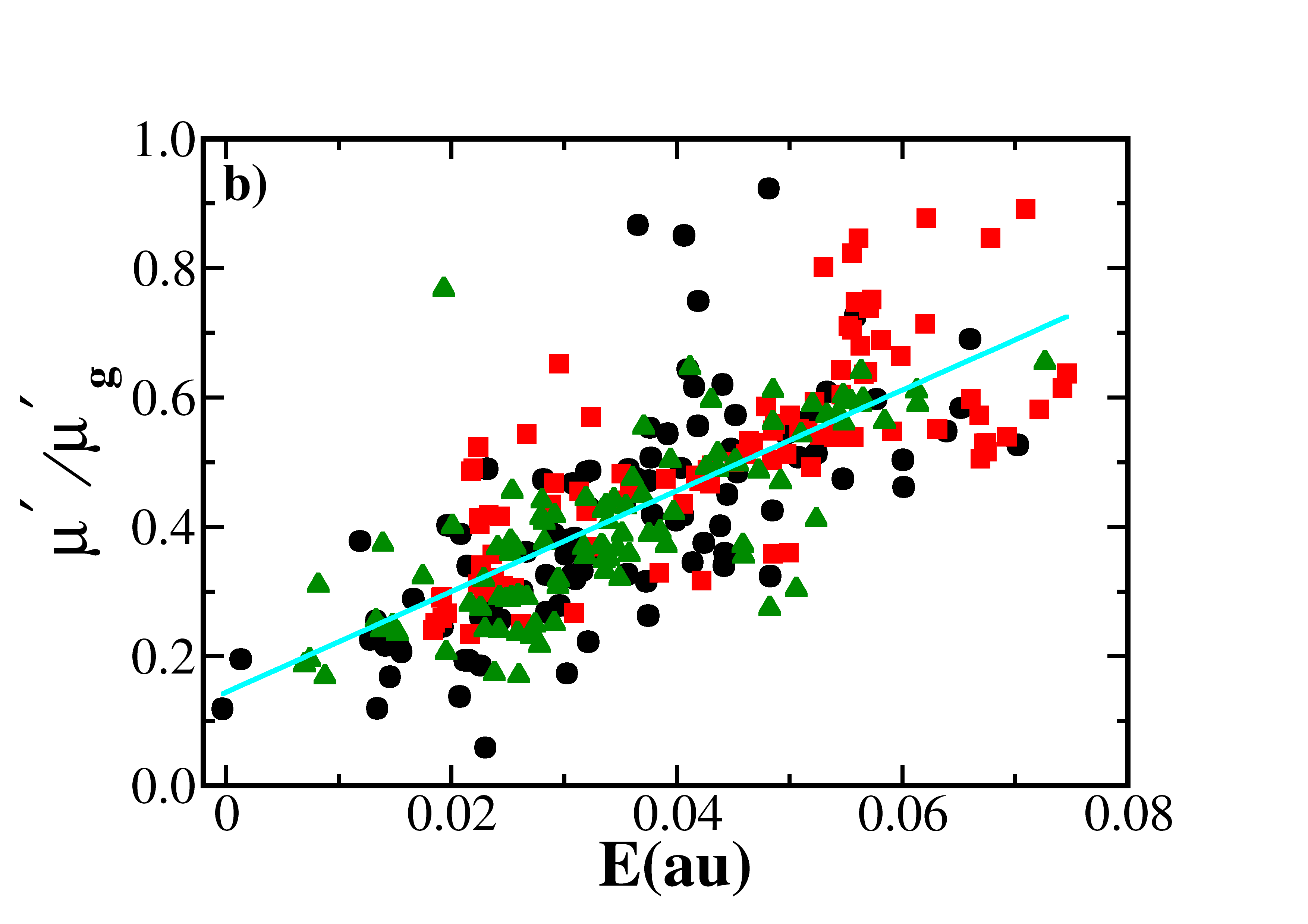}
    \includegraphics[width=0.45\textwidth, trim=0cm 0cm 2.5cm 1.5cm, clip]{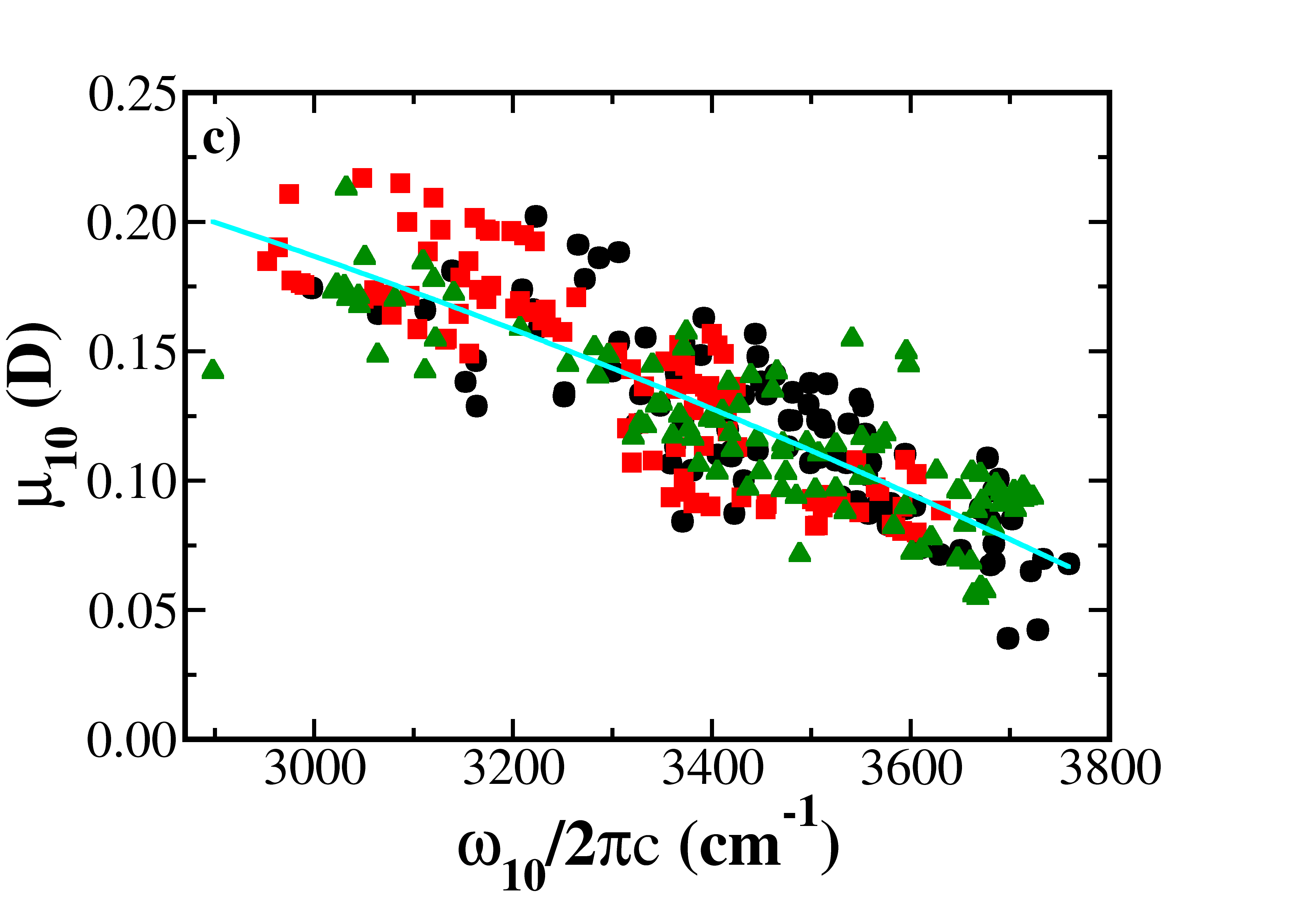}
    \caption{Correlations between \textbf{a)} the fundamental transition frequency and electric field (E) obtained from the DFT calculation, \textbf{b)} dipole moment derivative scaled by the gas phase dipole moment derivative, \textbf{c)} the fundamental transition frequency and transition dipole moment. All the black-colored symbols represent bulk water, and all other colored symbols represent urea-water data.} 
    \label{figure:MAPkbff}
\end{figure}

\begin{table}[htbp!]
    \centering
    \caption{Mapping Relation for KBFF - SPC/E Pair}
    \begin{tabular}{|c|}
        \hline
        \textbf{Mapping relations} \\
        \hline
        $\omega_{10} = 3788.18 - 9671.17E - 11394.5E^2$ \\
        $\mu' = 0.144465 + 7.78313E$ \\
        $\mu_{10} = 0.317041 + 4.81448 \times 10^{-5}\omega_{10} - 3.05235 \times 10^{-8}\omega_{10}^2$ \\
        \hline
    \end{tabular}
    \label{table:MAPkbff}
\end{table}

\newpage
\subsection{Maps for OPLS-S/OPLS-AA-D - SPC/E Pairs}
\begin{figure}[htbp!]
    \centering
    \includegraphics[width=0.45\textwidth, trim=0cm 0cm 2.5cm 1.5cm, clip]{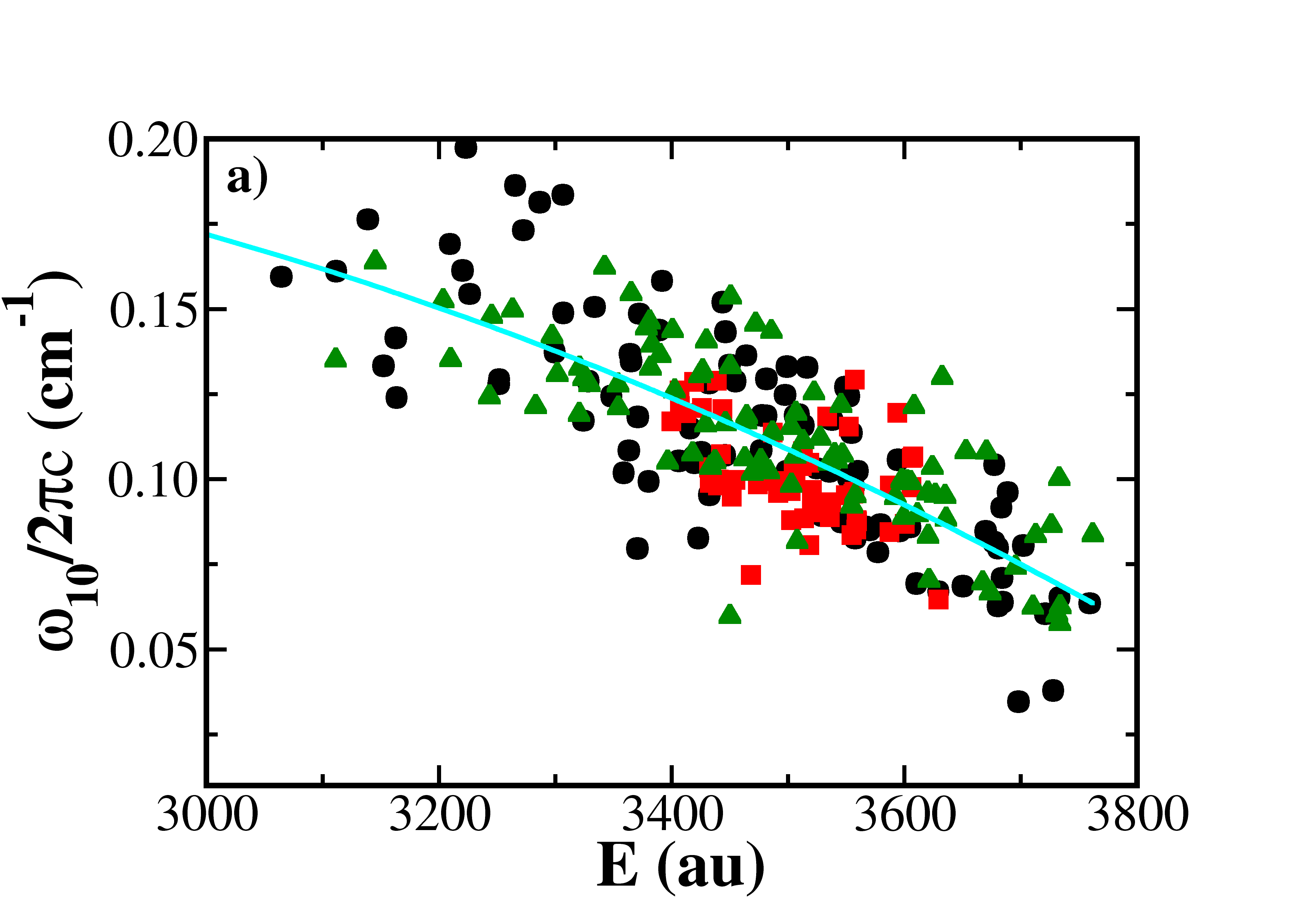} 
    \includegraphics[width=0.45\textwidth, trim=0cm 0cm 2.5cm 1.5cm, clip]{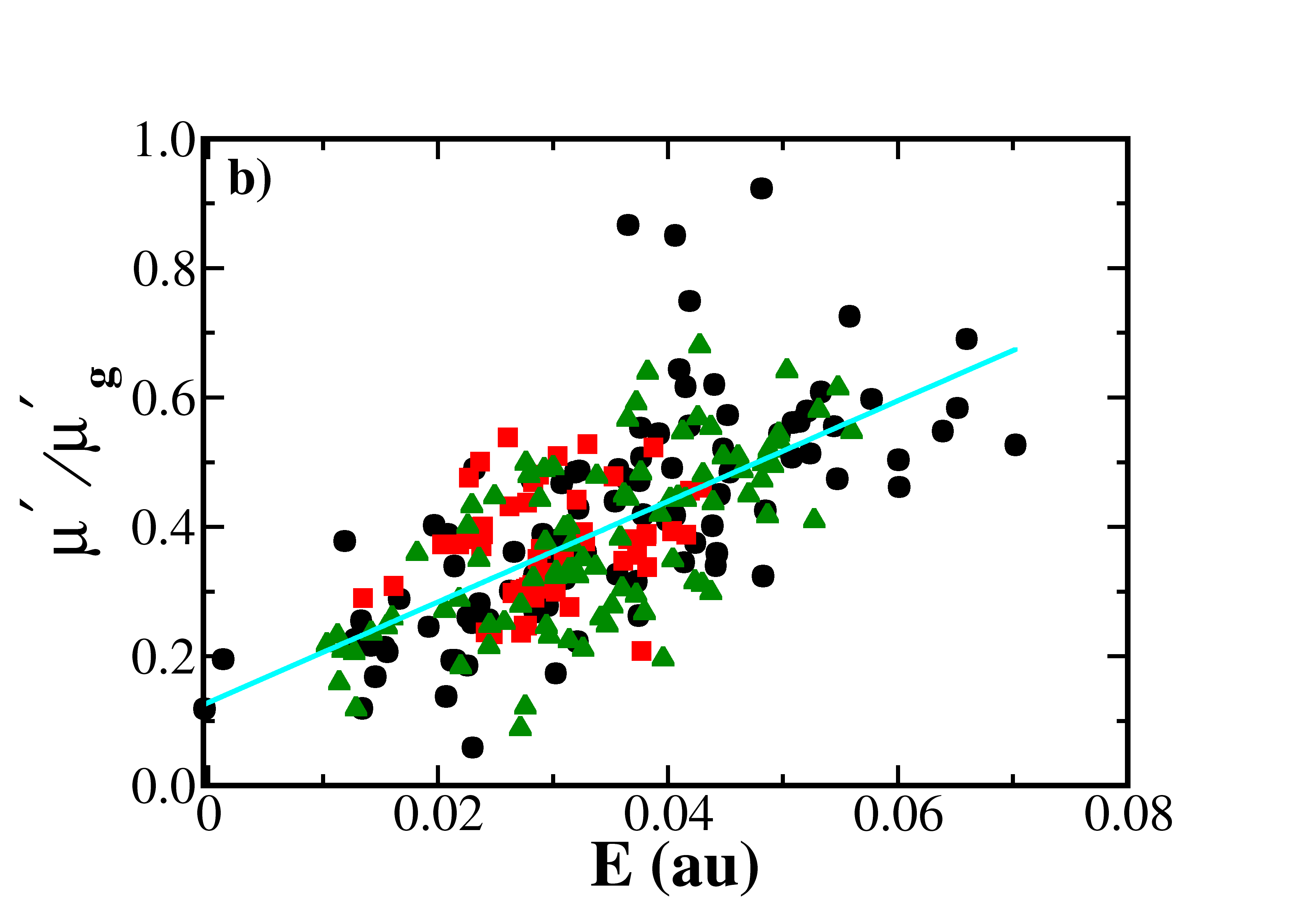}
    \includegraphics[width=0.45\textwidth, trim=0cm 0cm 2.5cm 1.5cm, clip]{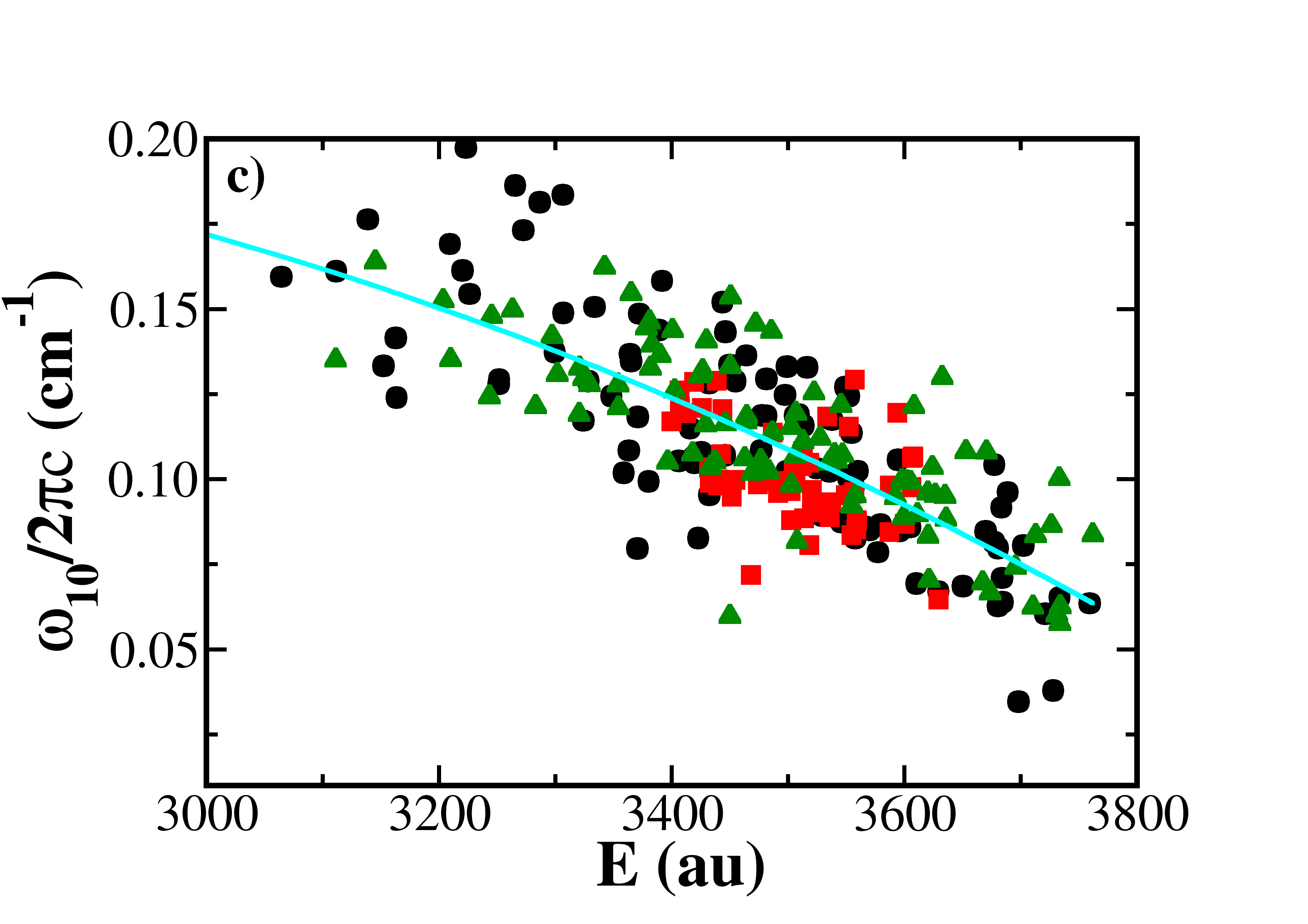}
    \caption{Correlations between \textbf{a)} the fundamental transition frequency and electric field (E) obtained from the DFT calculation, \textbf{b)} dipole moment derivative scaled by the gas phase dipole moment derivative, \textbf{c)} the fundamental transition frequency and transition dipole moment. All the black-colored symbols represent bulk water, and all other colored symbols represent urea-water data.} 
    \label{figure:MAPdfy}
\end{figure} 
\begin{table}[htbp!]
    \centering
    \caption{Mapping Relation for OPLS-S/OPLS-AA-D - SPC/E Pair}
    \begin{tabular}{|c|}
        \hline
        \textbf{Mapping relations} \\
        \hline
        $\omega_{10} = 3787.96 - 9247.33E + 4224.57E^2$ \\
        $\mu' = 0.12836 + 7.77615E$ \\
        $\mu_{10} = - 0.0818361 + 2.65529 \times 10^{-4}\omega_{10} - 6.03071 \times 10^{-8}\omega_{10}^2$ \\
        \hline
    \end{tabular}
    \label{table:MAPdfy}
\end{table}
\newpage
\subsection{Maps for GAFF-D3 - SPC/E Pair}
\begin{figure}[htbp!]
    \centering
    \includegraphics[width=0.49\textwidth, trim=0cm 0cm 2.5cm 1.5cm, clip]{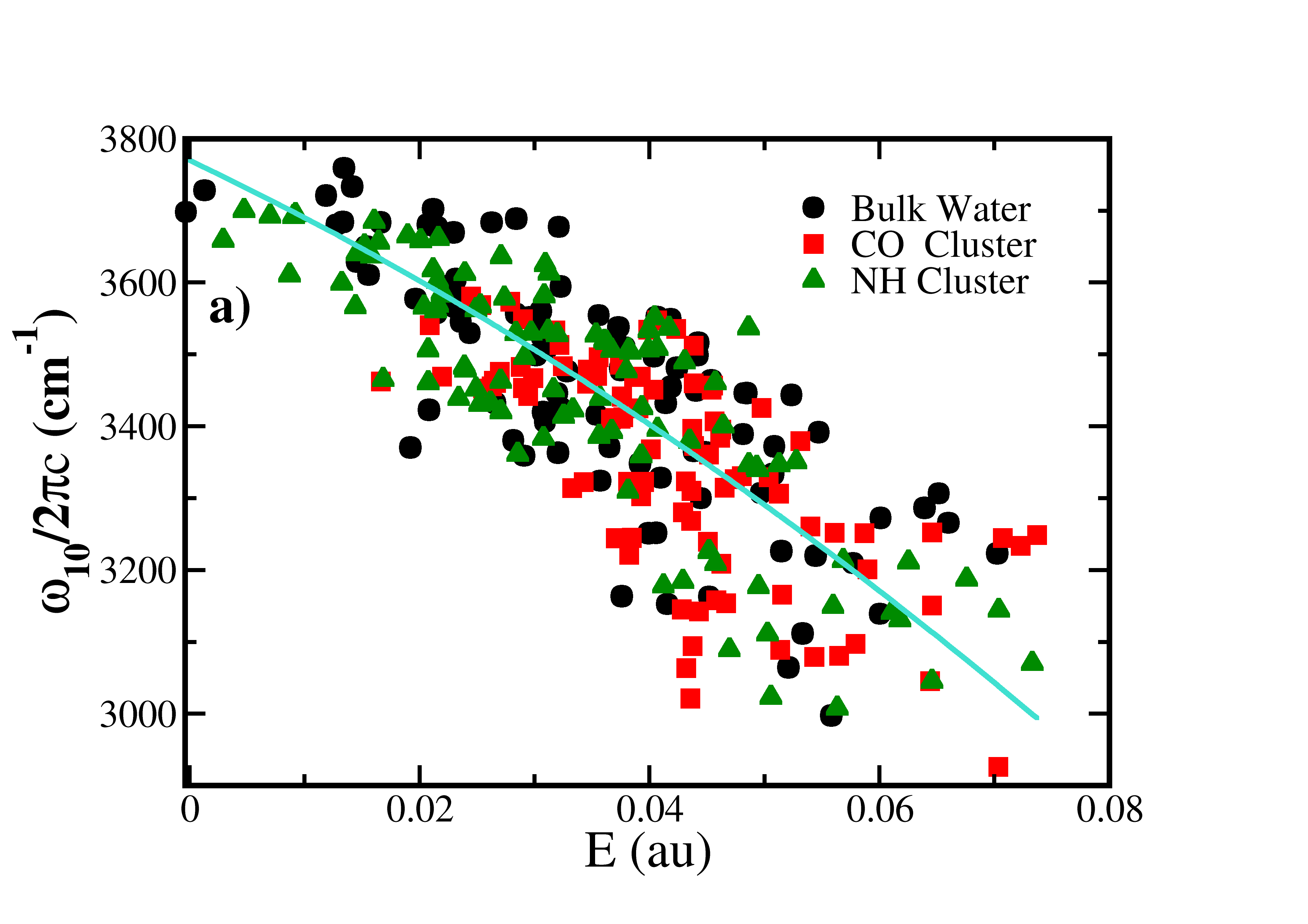} 
    \includegraphics[width=0.49\textwidth, trim=0cm 0cm 2.5cm 1.5cm, clip]{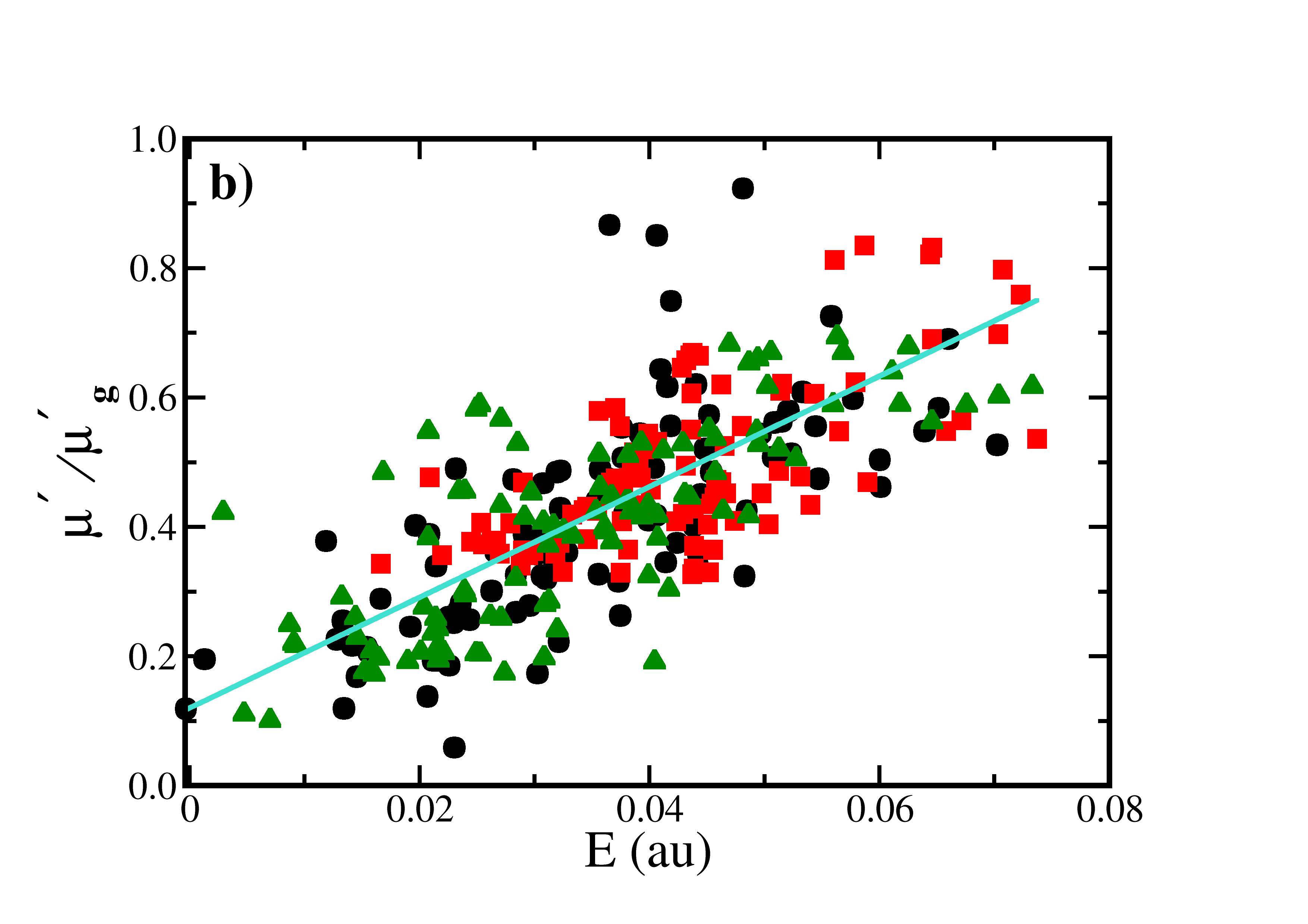}
    \includegraphics[width=0.49\textwidth, trim=0cm 0cm 2.5cm 1.5cm, clip]{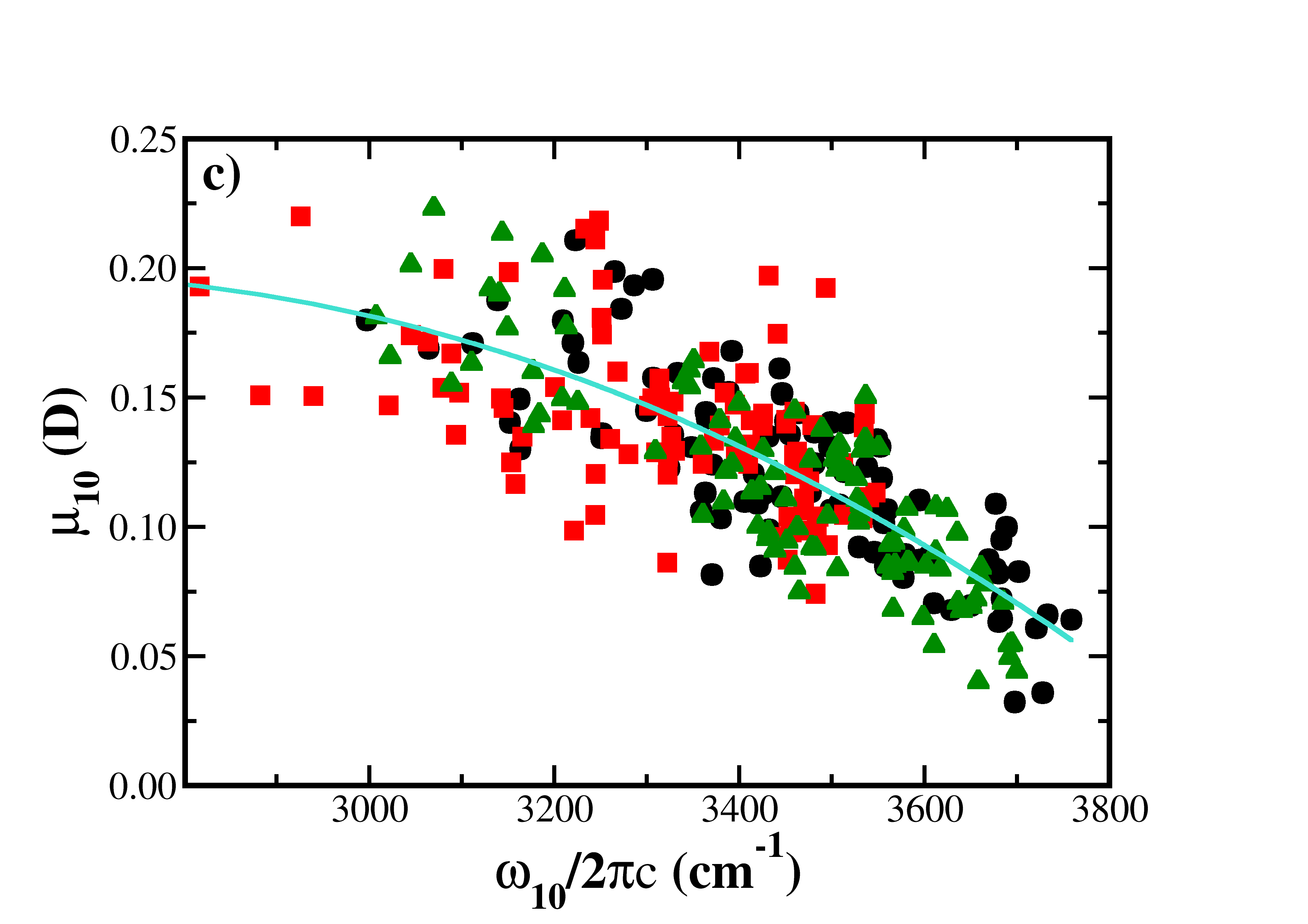}
    \caption{Correlations between \textbf{a)} the fundamental transition frequency and electric field (E) obtained from the DFT calculation, \textbf{b)} dipole moment derivative scaled by the gas phase dipole moment derivative, \textbf{c)} the fundamental transition frequency and transition dipole moment. All the black-colored symbols represent bulk water, and all other colored symbols represent urea-water data.} 
    \label{figure:MAPgfd3}
\end{figure} 
\begin{table}[htbp!]
    \centering
    \caption{Mapping Relation for GAFF-D3-SPC/E Pair}
    \begin{tabular}{|c|}
        \hline
        \textbf{Mapping relations} \\
        \hline
        $\omega_{10} = 3770 - 7587.15E - 39966.5E^2$ \\
        $\mu' = 0.12 + 8.55015E$ \\
        $\mu_{10} = - 0.549448 + 5.69872 \times 10^{-4}\omega_{10} - 1.08735 \times 10^{-7}\omega_{10}^2$ \\
        \hline
    \end{tabular}
    \label{table:MAPgaffd3}
\end{table}

\newpage
\section{Hydrogen Bond Strength}

We have used Harries and co-workers' recently proposed information theory-based scheme \cite{sapir2017revisiting} for the H-bond configurational analysis. Here is a brief explanation of the method. The truncated joint probability distribution $\widetilde{P}(r, \theta)$ is obtained using the two quantities, the donor-acceptor $(D-A)$ distance, r, and the hydrogen-donor-acceptor angle, $\theta$. Adjustment of the probability relative to the bulk is done by multiplying a weighting factor $F(r)$,
\begin{equation}
F(r) = \left\langle g(r)\frac{P_{rand}(r)}{\widetilde{P}(r)}\right\rangle
\end{equation}
where g(r) is the radial distribution function of the donor-acceptor pair, and the $P_{rand}(r)$ depends on the geometry of the pair of molecules that are considered. Therefore $P(r, \theta)$ is defined as follows,
\begin{equation}
P(r, \theta) = \widetilde{P}(r, \theta)\cdot F(r)
\end{equation}
Then, the probability is transformed to a free energy landscape of H-bond, 
\begin{equation}
PMF_{HB}(r,\theta) = - RT ln\frac{P(r, \theta)}{P_{rand}(r, \theta)}
\end{equation}
The H-bond strength is defined as the free energy quantified by the integral:
\begin{equation}
\Delta G = -RT \int\limits_{0}^{\pi} \int\limits_{r_{min}}^{r_{max}} \zeta P(r,\theta) \ln\frac{\zeta P(r, \theta)}{\zeta_{\text{rand}} P_{\text{rand}}(r,\theta)} \, d\theta \, dr \label{delG_eqsu},
\end{equation}
Here, the integration is done only over the volume of the first solvation shell, so the distribution functions are re-scaled by re-scaling factors $\zeta$ and $\zeta_{rand}$, respectively. Where ${1}/{\zeta}= \int\limits_{0}^{\pi} \int\limits_{r_{min}}^{r_{max}} P(r,\theta)d\theta\, dr $, and the expression for $\zeta_{rand}$ is analogous. $P_{rand}(r,\theta)$ is defined as  $P_{rand}(r,\theta)=P_{rand}(r) \cdot P_{rand}(\theta)$, where $P_{rand}(r)$ is the spherical shell distance degeneracy defined by,
\begin{equation}
P_{rand}(r) = \frac{\int_{r_i}^{r_{i+1}} 4\pi r^2 \, dr}{\int_{0}^{r_{max}} 4\pi r^2 \, dr}
\end{equation}
and $P_{rand}(\theta)$ is the random H-bond orientational distribution w.r.t $\theta$, which depends on the geometry of the pair of molecules considered. $P_{rand}(\theta)$ is calculated by generating random pair conformations of the molecules.

\section{Rotational Dynamics of Water With fitting}
Here we present C$_2$(t) vs time (ps) plots with fitting for pure water and urea-water binary solutions of different concentrations for all four forcefields of urea. 

\begin{figure}[htbp!]
    \centering
    \includegraphics[width=0.49\textwidth, clip]{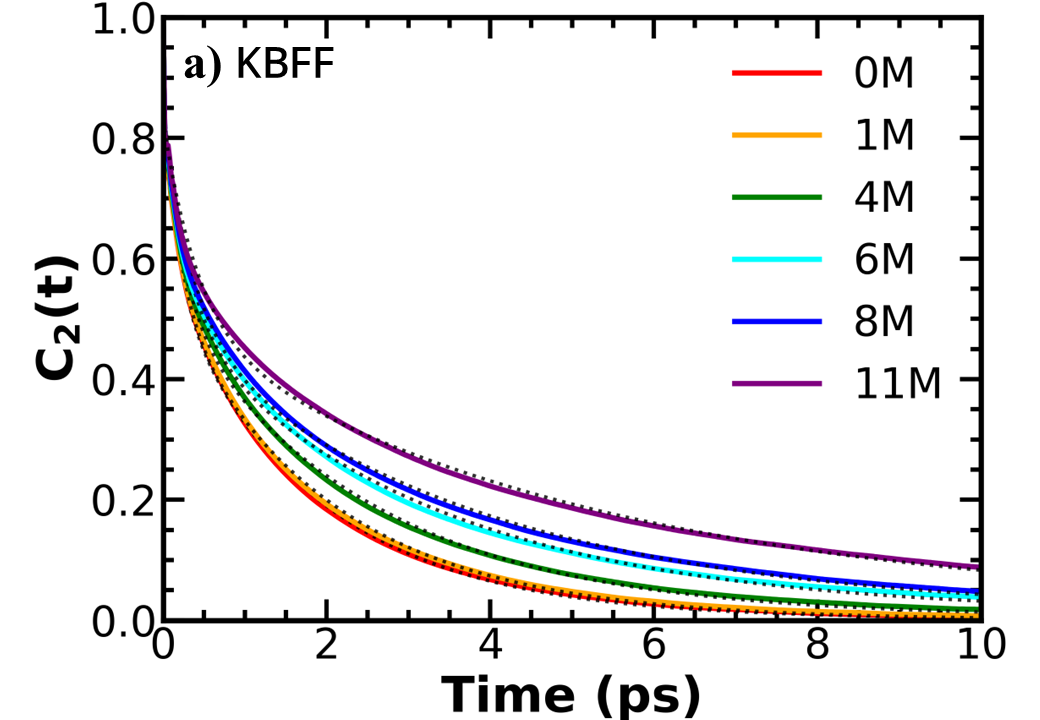} 
    \includegraphics[width=0.49\textwidth, clip]{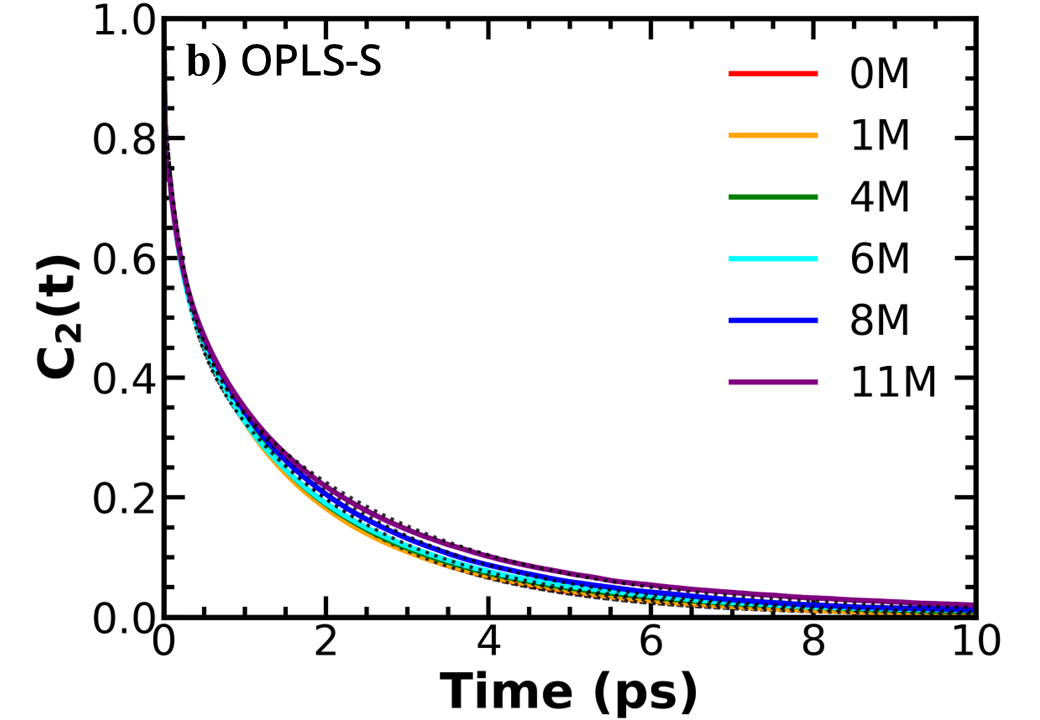}
    
        \vspace{0.25cm}  

    \includegraphics[width=0.49\textwidth, clip]{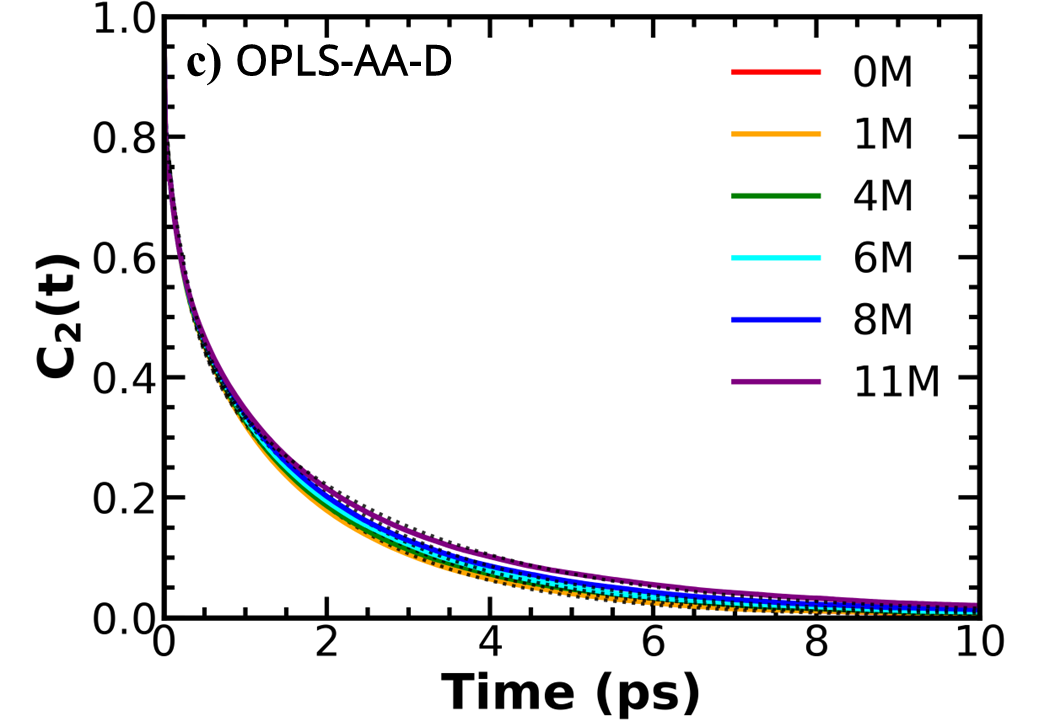}
    \includegraphics[width=0.49\textwidth, clip]{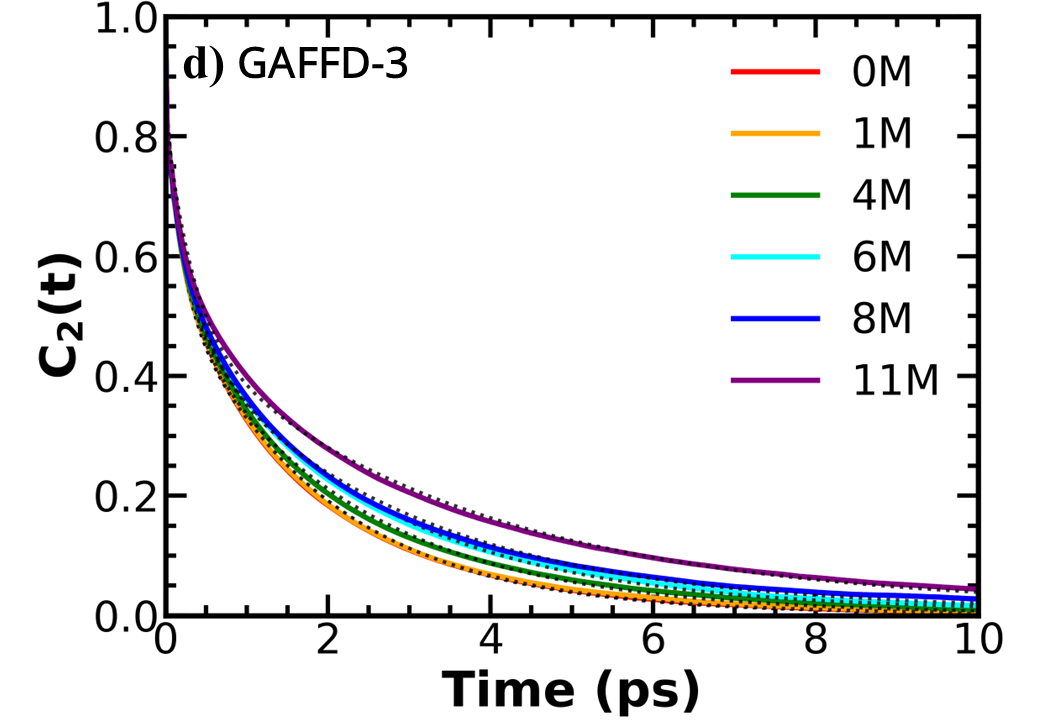}
    \caption {Orientational Time Correlation function (C$_2$(t)) of OH bond of pure water and various concentrations (1 M, 4 M, 6 M, 8 M, 11 M) of urea-water binary solution across different forcefields of urea with fitting, \textbf{a)} Kirkwood Buff Derived , \textbf{b)} OPLS-S , \textbf{c)} OPLS-AA-D , \textbf{d)} GAFF-D3. Black dotted lines represent a bi-exponential fitting curve.}
    \label{figure:fitc2t}
\end{figure}

\begin{figure}[htbp!]
    \centering
    \includegraphics[width=0.5\textwidth, trim= 1.5cm 1cm 2.5cm 2.5cm, clip]{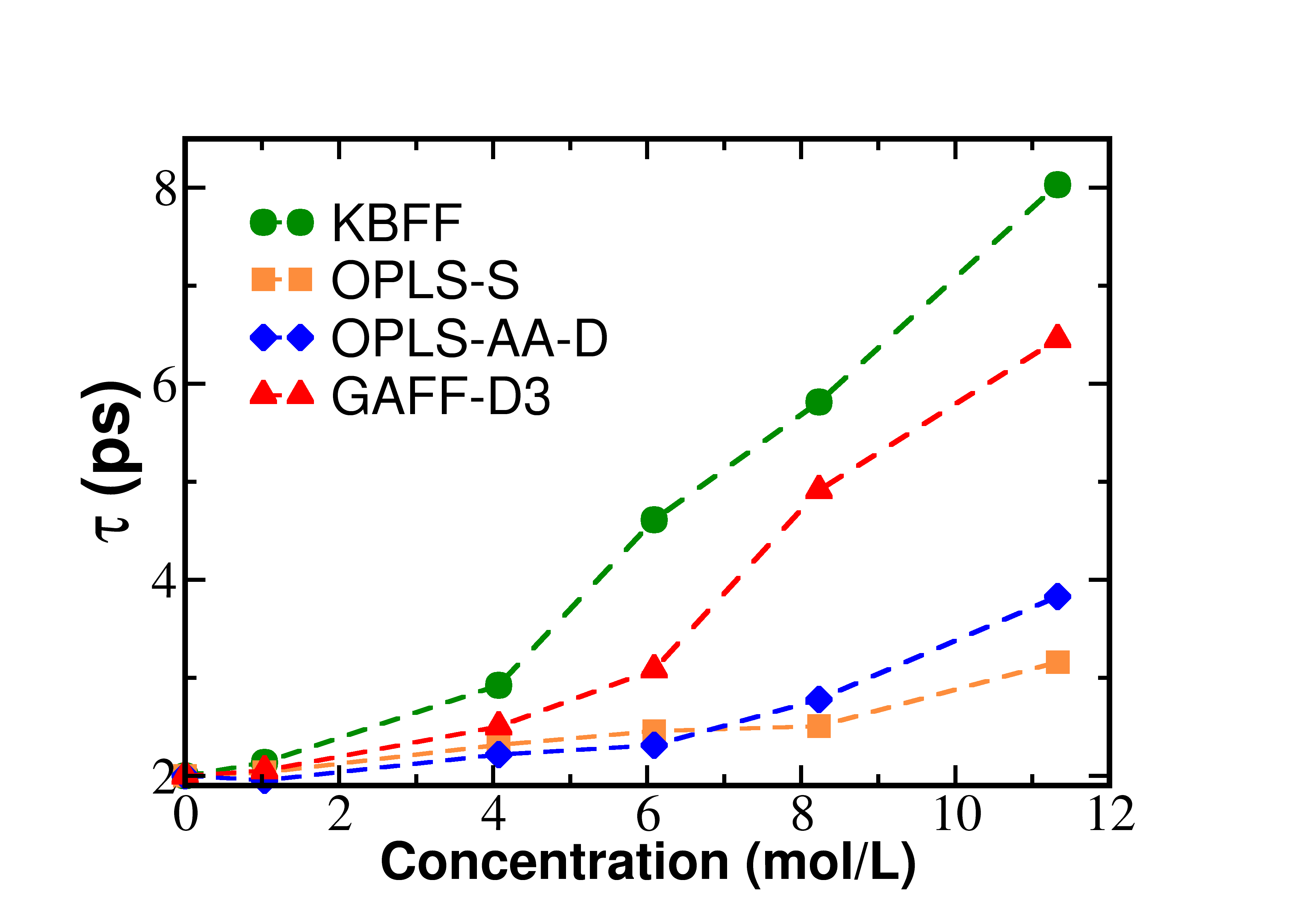} 
    \caption {The largest orientational relaxation time constant $\tau$ in ps as a function of urea concentration derived from multi-exponential fitting of C$_{2}$(t).}
    \label{figure:taumulti}
\end{figure}
Here is the multi-exponential fitting function containing a damp oscillation exponential to capture the librational motion of water\cite{moller2004hydrogen} as follows 
\[
y = a_0 e^{-x/\tau_1} + a_2 e^{-x/\tau} + a_4 e^{-x/\tau_2} \cos(a_6 x + a_7)
\]

\end{document}